\documentclass{article}
\usepackage[letterpaper,margin=1in]{geometry}
\usepackage{graphicx} % Required for inserting images
\usepackage{caption}
\usepackage{subcaption}
\usepackage{color} %% Required to use colors
\usepackage{hyperref} %% Required for using BibTeX
\usepackage{amssymb}
\usepackage[square,numbers]{natbib} 
\usepackage[table]{xcolor}
\usepackage{array}

\usepackage{orcidlink}

\newcolumntype{L}[1]{>{\raggedright\arraybackslash}p{#1}}
\newcommand\SgrA{Sgr~A$^{\star}$}

\begin{document}

%\maketitle

\begin{center}
    \vspace*{-1cm}
      \huge\textbf{Galactic Science with Ultra-High Angular Resolution X-ray Imaging}\\
      \Large\textbf{On Behalf of Working Group 3 of the Hi-ReX SAG}
    \vspace*{0.3cm}
  \end{center}  
    
    \large
    \noindent
    Paul A. Draghis\textsuperscript{1,*}\orcidlink{0000-0002-2218-2306}, 
    Jeremy Hare\textsuperscript{2,3,4,$\dagger$}\orcidlink{0000-0002-8548-482X}, 
    Mayura Balakrishnan\textsuperscript{5,6}\orcidlink{0000-0001-9641-6550},  
    Poshak Gandhi\textsuperscript{7}\orcidlink{0000-0003-3105-2615}, 
    Tyler Holland-Ashford\textsuperscript{2}\orcidlink{0000-0002-7643-0504}, 
    Margarita Karovska\textsuperscript{8}\orcidlink{0000-0003-1769-9201}, 
    Thomas Maccarone\textsuperscript{9}\orcidlink{0000-0003-0976-4755}, 
    Herman L. Marshall\textsuperscript{1}\orcidlink{0000-0002-6492-1293}, 
    Mark Reynolds\textsuperscript{10,11}\orcidlink{0000-0003-1621-9392}, 
    Malgosia Sobolewska\textsuperscript{8}\orcidlink{0000-0002-6286-0159}, 
    Ryan Tanner\textsuperscript{2,3}\orcidlink{0000-0002-1359-1626}
    \vspace*{0.3cm}
    
%author list: Paul A. Draghis, Jeremy Hare, Mayura Balakrishnan, Poshak Gandhi, Tyler Holland-Ashford, Margarita Karovska, Thomas Maccarone, Herman L. Marshall, Mark Reynolds, Malgosia Sobolewska, Ryan Tanner

     \normalsize   \noindent
    {\it Affiliations:} \textsuperscript{1}MIT Kavli Institute for Astrophysics and Space Research, Massachusetts Institute of Technology, 70 Vassar St, Cambridge, MA, 02139, USA\\
    \textsuperscript{2}NASA Goddard Space Flight Center, Astrophysics Science Division, Greenbelt, MD, 20771, USA\\
    \textsuperscript{3}The Catholic University of America, 620 Michigan Ave., N.E. Washington, DC, 20064, USA\\
    \textsuperscript{4}Center for Research and Exploration in Space Science and Technology, NASA/GSFC, Greenbelt, MD, 20771, USA\\
    \textsuperscript{5}Department of Physics, McGill University, 3600 Rue University, Montréal, Québec, H3A 2T8, Canada\\
    \textsuperscript{6}Trottier Space Institute at McGill, 3550 Rue University, Montréal, Québec, H3A 2A7, Canada\\
    \textsuperscript{7}School of Physics and Astronomy, University of Southampton, University Road, Southampton SO17 1BJ, UK\\
    \textsuperscript{8}Center for Astrophysics $\vert$ Harvard \& Smithsonian, 60 Garden St., Cambridge, MA, 02138, USA\\
    \textsuperscript{9}Department of Physics \& Astronomy, Texas Tech University, Box 41051, Lubbock, TX, 79409, USA\\
    \textsuperscript{10}Department of Astronomy, Ohio State University, 140 West 18th Ave., Columbus, OH, 43210, USA\\
    \textsuperscript{11}Center for Cosmology and AstroParticle Physics (CCAPP), Ohio State University, Columbus, OH, 43210, USA\\
    Corresponding authors email: \textsuperscript{*}pdraghis@mit.edu, \textsuperscript{$\dagger$}jeremy.hare@nasa.gov

\vspace*{0.2cm}

\begin{abstract}
\normalsize
\noindent Milli- to micro-arcsecond X-ray imaging will open a new observational regime for Galactic astrophysics by resolving physical scales that are inaccessible to current X-ray observatories. This white paper highlights the science enabled by such capabilities across four broad questions: how particles are accelerated, how stars die, how accretion is fueled, and what populations of X-ray sources inhabit the Galaxy. Ultra-high angular resolution will enable many new studies, such as resolving shocks and jets, measuring proper motions, parallaxes, and binary orbits, and providing secure multiwavelength counterpart identifications in crowded environments such as the Galactic Center and globular clusters. Combined with high-time-resolution observations, these measurements will connect variability and transient events to the physical structures in which they originate, while coordination with gravitational-wave, neutrino, $\gamma$-ray, radio, optical, and infrared facilities will provide spatial information needed to identify and characterize multi-messenger sources. In addition to defining the science cases, this paper presents a curated list of compelling targets spanning various angular resolutions and outlines the complementary specifications necessary to maximize the scientific outcome. Accomplishing these science goals will require an instrument with high angular resolution, precise astrometry, substantial collecting area, sufficient spectral and timing resolution, and high dynamic range imaging capabilities.
\end{abstract}

\section{Introduction}

X-ray observations provide a uniquely direct view of the hottest and most energetic components of astrophysical systems, including accretion flows onto compact objects, relativistic jets, shock-heated plasma, and populations of high-energy particles. However, many of the physical scales on which these processes originate remain far below the angular resolution of current X-ray observatories. As a result, phenomena that are spatially resolved at optical, infrared, or radio wavelengths are often observed in X-rays only through their integrated spectra and variability. Advancing X-ray imaging from the sub-arcsecond regime into the milli- to micro-arcsecond regime would therefore provide spatial information to problems that are currently constrained primarily through indirect measurements, allowing the locations, geometries, and motions of X-ray-emitting structures to be measured directly.

The high angular resolution study of Galactic sources would unlock unique insights into the fundamental mechanisms that govern the Universe. The Milky Way provides an especially powerful laboratory as its proximity allows physical scales ranging from supernova remnant shocks to compact-binary separations, and to the environments in the immediate vicinity of black holes, to be mapped at angular scales that would remain inaccessible in extragalactic systems. At the same time, precise X-ray astrometry would extend these capabilities beyond imaging alone, enabling measurements of proper motions, parallaxes, and orbital motion for sources that may be obscured, crowded, or undetectable at other wavelengths. In particular, this work highlights how improved imaging and astrometric capabilities can address four broad questions:

\begin{enumerate}
\item How are particles accelerated? (Section \ref{sec:part_accel})
\item How do stars die? (Section \ref{sec:stellar_death})
\item How is accretion fueled? (Section \ref{sec:accretion_fuel})
\item What are the populations of X-ray sources? (Section \ref{sec:xray_pop})
\end{enumerate}

\begin{figure}[ht]
    \centering
    \includegraphics[width=1.0\linewidth]{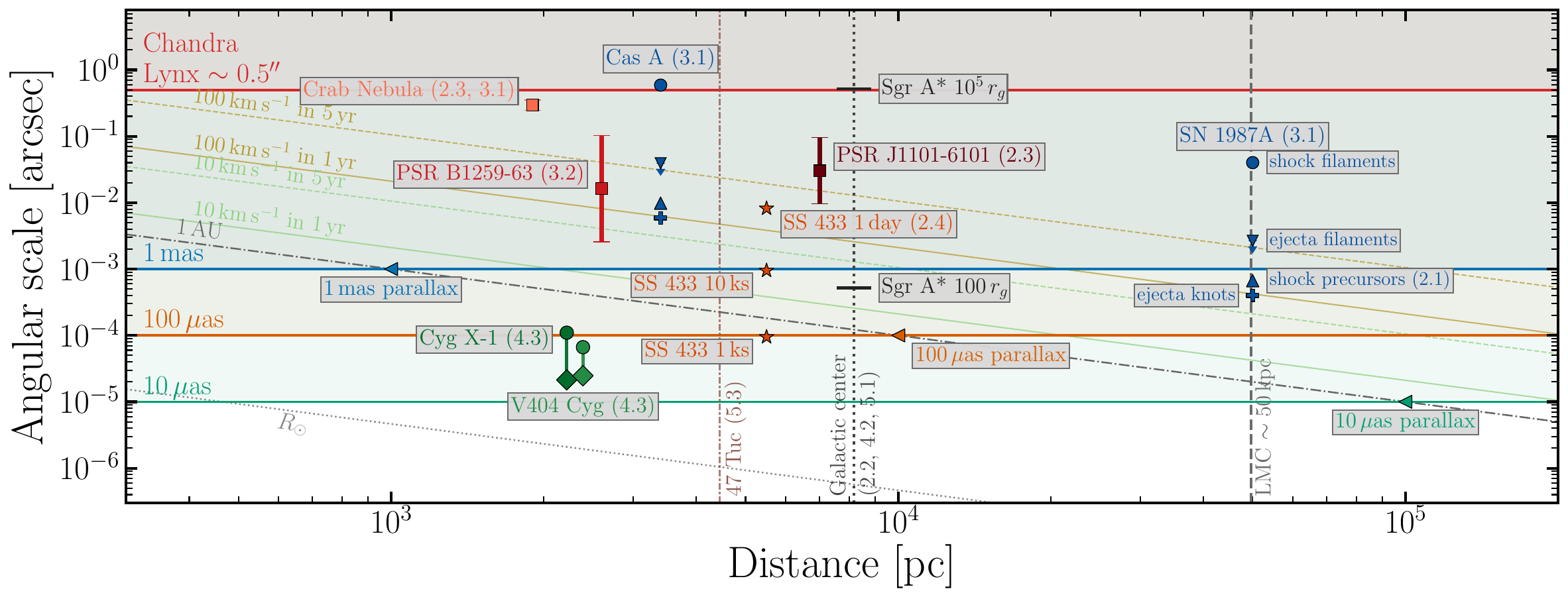}
    \caption{Representative physical and angular scales for the Galactic science cases discussed in this work. Source distance is shown along the horizontal axis and angular scale along the vertical axis, with horizontal bands marking the approximate resolution regimes of 1~$m$as, 100~$\mu$as, and 10~$\mu$as. Individual markers indicate characteristic scales for selected sources and phenomena, including compact-binary separations, supernova-remnant structures, pulsar-wind features, jet ejecta, Galactic-center accretion scales, and astrometric measurements such as parallaxes and proper motions. The figure illustrates how progressively finer angular resolution opens access to qualitatively different science, from resolving extended shocks and ejecta to probing compact-binary and black-hole environments directly. The numbers next to certain markers indicate the sections where these science cases are discussed. }
    \label{fig:resolution_limits}
\end{figure}

Figure \ref{fig:resolution_limits} displays the broad range of physical and angular scales relevant to these science cases. The horizontal bands indicate representative angular-resolution milestones of 1~$m$as, 100~$\mu$as, and 10~$\mu$as, illustrating that different stages in the development of ultra-high-resolution X-ray imaging would already open distinct regions of discovery space. Rather than requiring the ultimate angular resolution for every application, many of the science cases considered below become progressively accessible as the resolution improves. The following sections connect these angular scales to specific Galactic sources and physical processes, demonstrating both the breadth of the science enabled and the complementary requirements on astrometric precision, collecting area, energy and time resolution, and imaging dynamic range.

\section{How are Particles Accelerated?} 
\label{sec:part_accel}

\subsection{Supernova Remnants (SNRs)} 
\label{sec:part_accel_SNR}

SNRs are theorized to be responsible for cosmic ray acceleration up to the ``knee'' at $\sim$3 PeV (see e.g., \cite{2012A&ARv..20...49V} for a review). In order for these particles to be accelerated, the ambient ISM magnetic field of $\sim$3~$\mu$G must be amplified more than the factor of $\lesssim$4 from shock compression. Studies of the thin synchrotron-emitting filaments along the edges of SNRs have supported magnetic field amplification up to 50--500~$\mu$G in MW SNRs. The thin widths of these filaments necessitates a rapid dropoff in downstream synchrotron emission from either electron energy losses or magnetic field damping, and the diffusive shock acceleration process requires a precursor shock from upstream accelerated particles rebounding back back towards the shock front. However, evidence of these phenomena and their origins have yet to be conclusively found.

\begin{figure}[hb]
\centering
\includegraphics[width=0.32\linewidth]{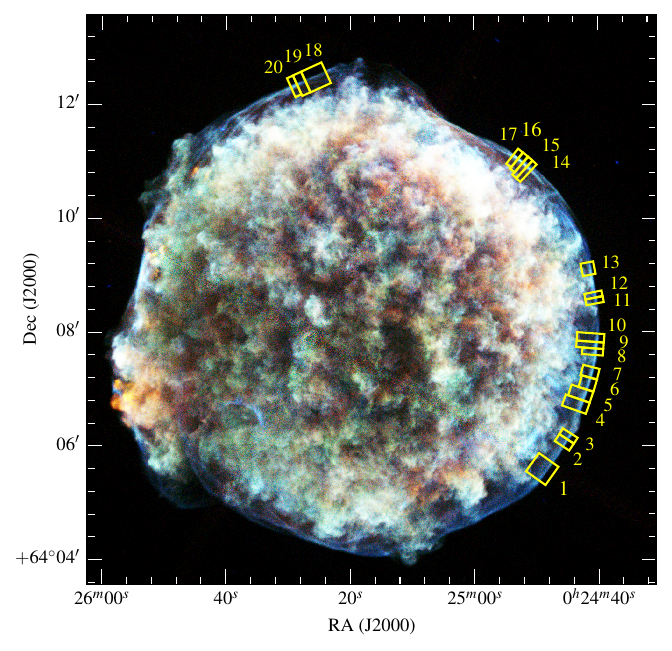}
\includegraphics[width=0.32\linewidth]{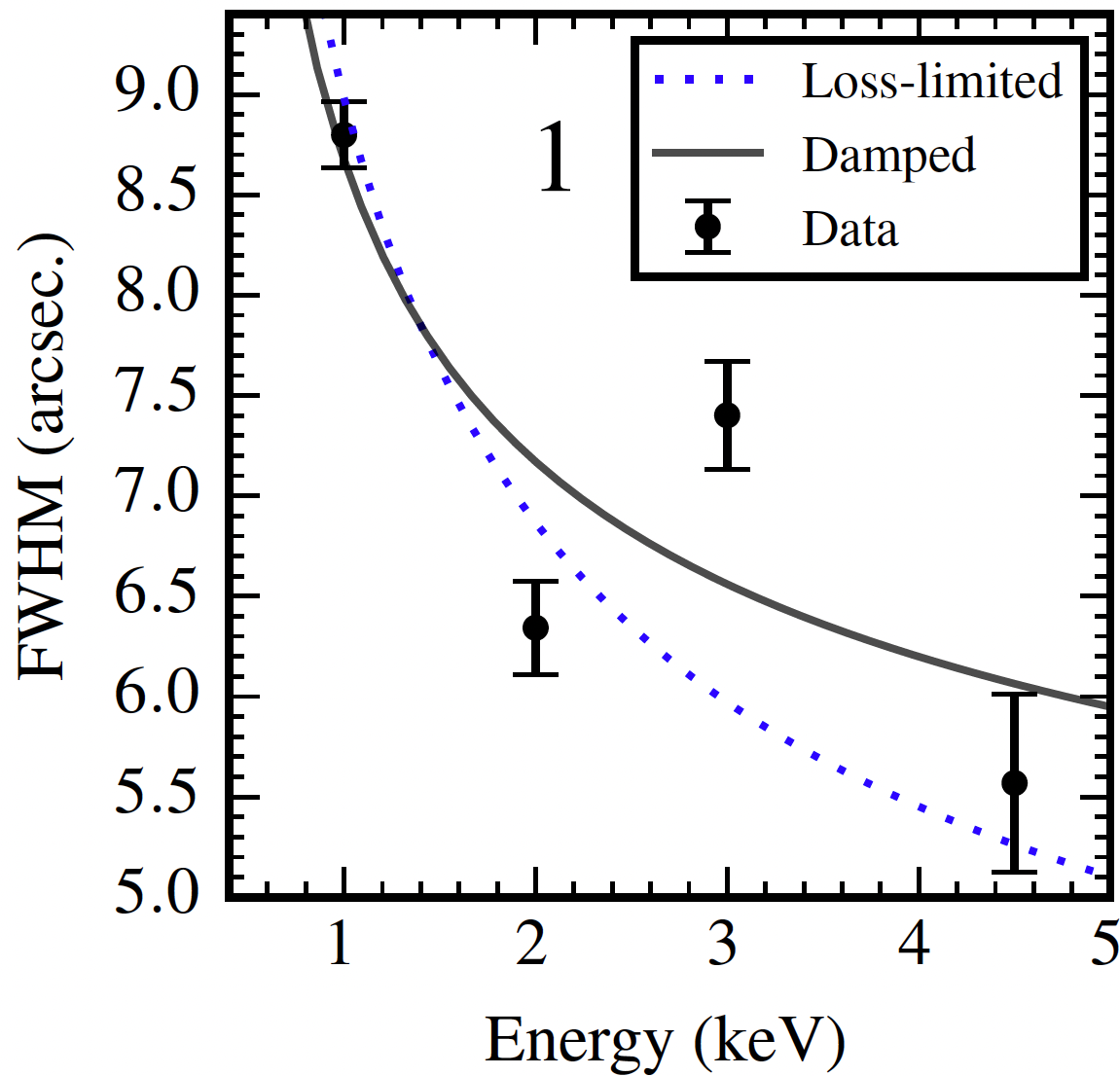}
\includegraphics[width=0.32\linewidth]{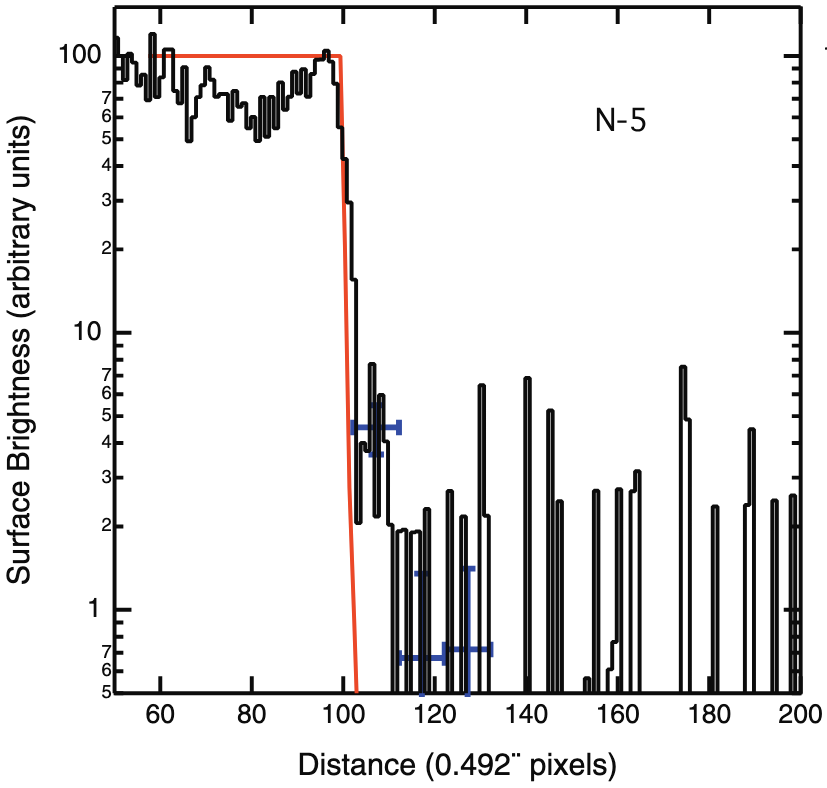}
\caption{ (Left): Chandra RGB observation of Tycho's SNR labeled with filament regions selected for energy-width study. (Middle): the data from region 1 fit with both loss-limited and magnetic damping models, taken from \cite{2015ApJ...812..101T}. Neither model accurately fits the data. (Right): the observed profile of the upstream shock for a region in SN~1006 (black) which cannot be statistically distinguished from the dropoff due to the Chandra PSF response to a sharp edge (red), taken from \cite{2014ApJ...781...65W}. Resolving these structures at one to two orders of magnitude below Chandra would allow direct measurements of both the downstream synchrotron-loss scale and any upstream precursor emission, providing substantially stronger constraints on magnetic-field amplification and particle diffusion at SNR shocks. }
\label{fig:tycho_rims} 
\end{figure}

\textbf{Downstream:} \cite{2014ApJ...790...85R} found that filament widths in SN~1006 depended on photon energy, ruling out magnetic damping models. However, \cite{2015ApJ...812..101T} found evidence consistent with either loss-limited or magnetic damping models in Tycho's SNR (see Figure~\ref{fig:tycho_rims}). Additionally, the magnetic damping models allowed much lower magnetic fields immediately downstream of the shock: only $\sim$10$\times$ that of typical galactic field values. As typical SNR filament widths are 0.01--0.1~pc upstream and 0.06--0.4~pc downstream (corresponding to $\sim$0.5''--20'' for Galactic SNRs; \cite{2005ApJ...621..793B}), a moderate angular resolution of $\lesssim$0.1'' will be sufficient to robustly measure the filament width dependence on energy and explain the rapid dropoff in downstream synchrotron emission in SNRs.

\textbf{Upstream:} Diffusive shock acceleration requires that accelerated particles spend time ahead of the shock (before they bounce back toward the shock on the turbulent magnetic field lines) and produce synchrotron radiation while there. However, evidence of this has yet to be conclusively discovered at levels below Chandra's PSF (e.g., \cite{2014ApJ...781...65W}; see Figure~\ref{fig:tycho_rims}, Right). Estimates of the scale length for pre-shock emission are bounded by the electron gyroradius and the diffusive scale length. For typical SNR magnetic field strengths of 50--500 $\mu$G, the electron gyroradius for a particle radiating a peak of 4~keV is (1.3--42) $\times \ 10^{14}$ cm and the Bohm diffusive scale length is (2.7--85) $\times \ 10^{15}$ cm. Using a conservative intermediate value of 1.3 $\times 10^{15}$ cm, a telescope with an angular resolution of $\lesssim$0.01'' is needed to identify shock precursors for SNRs 8~kpc away (0.025'' for D=3.4~kpc).However, the findings of \cite{2014ApJ...790...85R} suggest the possibility of ``sub-Bohm diffusion'' where the mean free path can be less than the gyroradius in perpendicular shocks. This might necessitate even smaller angular resolution X-ray telescopes, going down to $\sim$1~$m$as.

\subsection{Sgr A*: Jets and Outflows}
\label{sec:sgrA_jets_out}

\begin{figure}[!hb]
\centering
\includegraphics[width=0.7\linewidth]{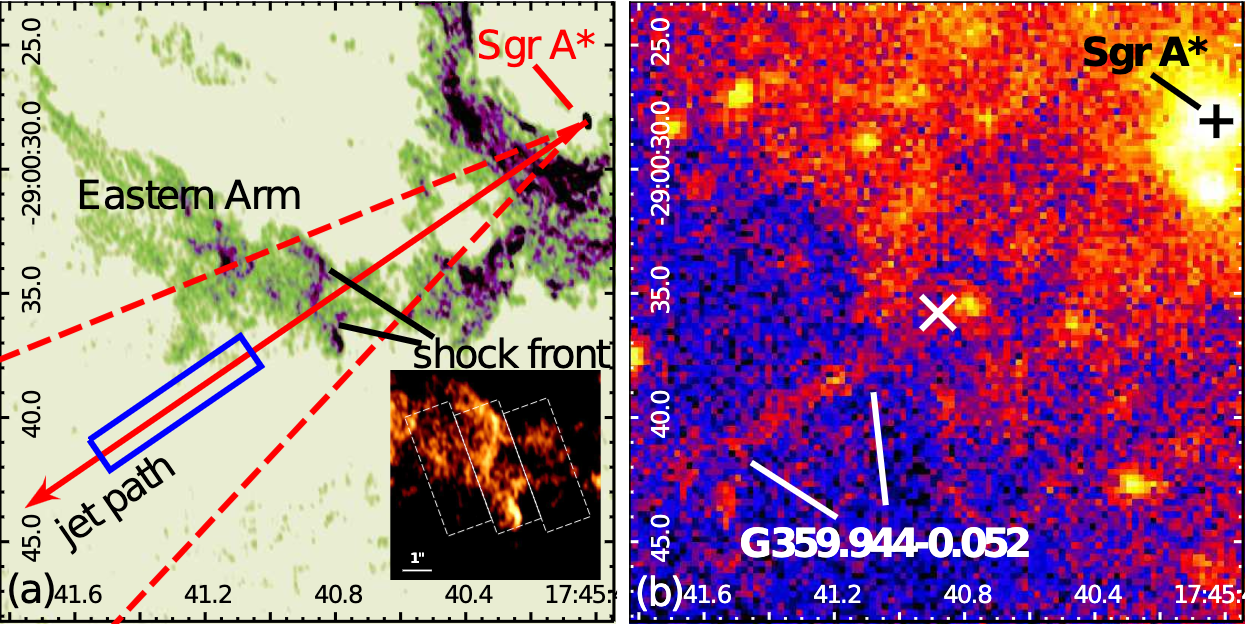}
\caption{Multiwavelength evidence for a candidate parsec-scale jet from Sgr~A$^*$. Panel (a) shows the VLA 1.3-cm emission from the Eastern Arm of Sgr~A West, including a possible shock front and the proposed jet path from Sgr~A$^*$. Panel (b) shows the corresponding Chandra 2--8 keV image, in which the linear filament G359.944$-$0.052 lies downstream of the shock. This alignment has been interpreted as a jet interacting with the surrounding ionized gas, although its association with Sgr~A$^*$ remains unconfirmed. Milli-arcsecond X-ray imaging could test whether a collimated inner outflow connects Sgr~A$^*$ to this larger-scale structure. Reproduced from \cite{Li2013}.}
\label{fig:sgra_candidate_jet}
\end{figure}

Jets and outflows provide direct laboratories for studying particle acceleration in the immediate environments of accreting black holes, where magnetic energy can be converted into bulk motion and non-thermal particles. As the closest supermassive black hole (SMBH), Sagittarius A$^*$ (Sgr A$^*$) offers a unique opportunity to spatially resolve these particle-acceleration processes on scales that are inaccessible in more distant systems. Deep Chandra spectroscopy provides evidence for mass loss in an outflow from the radiatively inefficient accretion flow, although no collimated jet has yet been detected directly \cite{Wang2013}. A linear X-ray filament, G359.944$-$0.052, and an associated radio shock front have been interpreted as a candidate parsec-scale jet, but these features do not resolve the jet-launching region \cite{Li2013,Zhu2019}. As shown in Figure \ref{fig:sgra_candidate_jet}, the X-ray filament lies downstream of the proposed radio shock and approximately along a possible jet trajectory from Sgr A$^*$, although the physical association remains unconfirmed. A key goal of $m$as X-ray imaging would therefore be to determine whether an inner collimated outflow connects Sgr A$^*$ to this much larger-scale structure.

At a distance of 8.1~kpc and a mass of $4.1\times10^6\,M_\odot$ \cite{GRAVITY2019}, 1~mas corresponds to approximately $200\,r_g$. This scale probes an important gap between the horizon-scale structure measured by the Event Horizon Telescope and the much larger-scale candidate outflow seen in Figure \ref{fig:sgra_candidate_jet}. EHT polarimetry has revealed ordered magnetic fields close to the event horizon, providing conditions favorable for magnetically driven outflows \cite{EHT_VII,EHT_VIII}, while mm-VLBI studies indicate that projected scales of $\sim0.4$--2.5~mas ($\sim70$--$460\,r_g$) may encompass the region in which a compact outflow becomes collimated \cite{Park2015}. Imaging Sgr A$^*$ at $\sim1$~mas could therefore test whether the X-ray core remains point-like or develops a persistent elongation, knots, or other asymmetric structure associated with a weak jet or episodic magnetized outflow. The best-fitting 7-mm jet models favor a bipolar flow viewed at high inclination and tentatively suggest a sky position angle near $105^\circ$, although these constraints remain model dependent \cite{Markoff2007}. High-resolution X-ray imaging could directly test whether any resolved structure preferentially develops along this axis and could probe the region where magnetic energy is converted into bulk kinetic energy and accelerated particles.

Variability provides an additional dimension to this experiment. Sgr~A$^*$ also displays significant variability across the electromagnetic spectrum with X-ray flares reaching $\sim10^2$--$10^3$ times the quiescent level (on average $\sim$1 per day; \cite{Neilsen2013}) and NIR flux variations of factors of a few to $\sim$100 ($\sim$4 per day; \cite{Schodel2011,Do2019}). Milli-arcsecond imaging could determine whether the morphology of the X-ray source changes during these events, for example through the appearance of transient elongation, moving ejecta, or localized particle-acceleration sites. Such measurements would directly connect temporal variability to changes in spatial structure. For this science case, a field of view of at least $1''\times1''$ is required. Given the quiescent \textit{Chandra} count rate of $5.2\times10^{-3}$ counts s$^{-1}$ \cite{Neilsen2013}, an effective area of approximately $3000$~cm$^2$ in the 4--8~keV band would provide $\sim3000$ counts in 100~ks, enabling flare-resolved imaging and spectroscopy. An energy resolution of $\sim100$~eV at 6~keV would be sufficient for broad morphological spectroscopy, while $\sim6$--7~eV resolution would enable kinematic measurements; high-time-resolution capability would also be required to follow the rapidly evolving flares.

\subsection{Magnetars, Pulsar Wind Nebulae, and PeV Sources} 
\label{sec:mag_psr_PeV}
Pulsars lose their rotational energy primarily through relativistic magnetized particle winds known as pulsar wind nebulae (PWN; see e.g., \cite{2017hsn..book.2159S} for a review). The relativistic wind from the pulsar eventually collides with the surrounding material (i.e., either the material in the SNR, or ISM once the pulsar has escaped its SNR) and is slowed down at the termination shock (TS), where the energetic particles interact with the magnetic field and produce synchrotron and inverse Compton emission. The size of the TS is determined by the distance at which the pressure from the pulsar wind equals the pressure from the surrounding medium, D$_{\rm TS}\approx(\dot{E}/4\pi c P_{\rm amb})^{1/2}\approx0.05\dot{E}_{\rm{36}}^{1/2}P^{-1/2}_{\rm amb,10}$ pc, where $\dot{E_{\rm{36}}}$ is the pulsar spin-down power in units of 10$^{36}$ erg s$^{-1}$ and $P_{\rm amb,10}$ is the ambient pressure in units of 10$^{-10}$ dyn cm$^{-2}$ (scaling adopted from \cite{2008AIPC..983..171K}). While the termination shock distance is typically large for young, energetic pulsars (e.g., $\sim1'$ for the Crab pulsar), substructure has been observed on scales much smaller than the TS. For instance, knots have been observed in the Crab PWN on scales as small as several hundred $m$as for the knots in X-rays, optical, and radio \cite{2002ApJ...577L..49H,2011Sci...331..736T,2011A&A...533A..10L,2015ApJ...811...24R}. Probing these much smaller scales in X-rays would allow for searches of X-ray substructure within the knots and wisps, which can help further inform broadband modeling of the PWN. Additionally, there are several pulsars that are relatively bright in X-rays, making it difficult to resolve their PWN due to the bright PSF wings that overlap it (see, e.g., \cite{2009ApJ...690..891K,2024ApJ...968...67G}). High angular resolution imaging would help to firmly separate the PWN emission from that of the pulsar.

\begin{figure}[!b]
    \centering
    \includegraphics[width=1.0\linewidth]{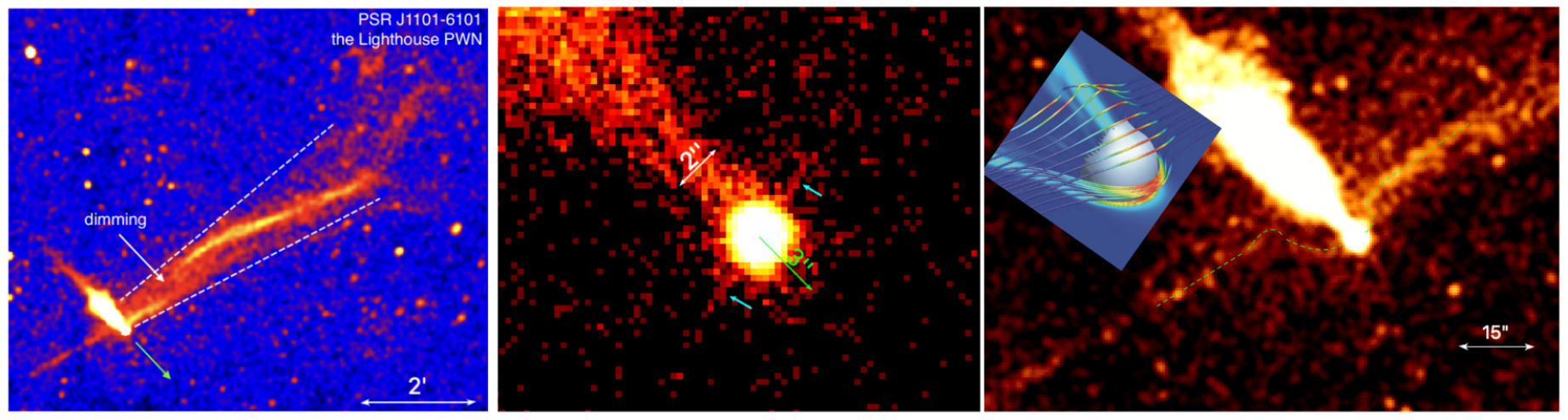}
    \caption{Chandra images of PSR J1101–6101. The left panel shows a zoomed out image of PSR J1101–6101's pulsar wind nebula and long mis-aligned outflow. The second image shows the protrusions near the head of the PWN extending from the point source (cyan arrows) on arcsecond scales. The last panel shows a model of magnetic draping of the ISM magnetic field lines near the pulsar wind bow shock. Figure adopted from \cite{2023ApJ...950..177K} and modified.}
    \label{fig:lighthouse}
\end{figure}

Another interesting facet of PWN science are the spectacular mis-aligned outflows. One of the most prominent examples is the mis-aligned outflow from the lighthouse PWN (see Figure \ref{fig:lighthouse} and \cite{2023ApJ...950..177K}). These extended X-ray emitting structures are thought to be formed by the highest energy PWN particles that escape from the apex of the PWN bow-shock and travel along the ISM magnetic field lines, producing synchrotron emission. However, several other models have also been suggested (see e.g., \cite{2023Univ....9..402O,2024ApJ...976....4D}). In any case, high angular resolution imaging may help shed light on exactly how the particles escape into the ISM, by resolving the structures near the PWN bow-shock ($\sim$1-100 $m$as angular scales).

Magnetars are neutron stars with high magnetic fields (i.e., $B\gtrsim10^{13}$ G), which exhibit strong X-ray outbursts. On rare occasion, they can produce giant flares, with the most energetic flares having a total energy of $E\gtrsim10^{46}$ erg and reaching an X-ray luminosities of $L_X\gtrsim10^{47}$ erg s$^{-1}$ (e.g., SGR 1806-20; \cite{2005Natur.434.1098H}). These flares are so energetic that they can launch material from the NS surface having velocities of $\sim0.25c$ \cite{2005ApJ...634L..93T}. Radio observations of SGR J1806-20 found asymmetric resolved radio emission, having a size of 50-300 $m$as, around the magnetar about 10 days after its giant flare in 2004 \cite{2005ApJ...634L..93T}, which showed signs of deceleration at late times. Magnetars are also leading candidates for the origin of at least some fast radio bursts (FRBs), a connection established by the detection of the Galactic FRB 200428 from the magnetar SGR 1935+2154, accompanied by an X-ray burst (\cite{2020Natur.587...54C,2020Natur.587...59B}). Resolving the X-ray-emitting ejecta and surrounding environment following magnetar flares could therefore help connect the conditions associated with magnetar activity to the production of coherent radio bursts, providing an important nearby laboratory for understanding the FRB phenomenon. Furthermore, it has been suggested that these giant flares may be appreciable sites of heavy r-process nucleosynthesis \cite{2024MNRAS.528.5323C}. A high-resolution X-ray observatory would allow for such ejecta to be monitored in X-rays at early times to constrain the particle injection spectrum and cooling and help constrain broadband models of the ejecta. This could also help to better differentiate between extragalactic magnetar giant flares and GRBs \cite{2021ApJ...907L..28B,2024Natur.629...58M,2026A&A...710A.139T}. Lastly, on the fainter end, at least one magnetar (Swift J1834.9-0846) has shown extended X-ray emission several years after undergoing an outburst \cite{2016ApJ...824..138Y}, which could be studied with a high angular resolution X-ray observatory.

High energy (TeV) observatories have uncovered a large population of TeV emitting PWN \cite{2018A&A...612A...2H}, some of which may accelerate electrons accelerated to PeV energies (see e.g., \citealt{2026ApJ...998..230B}). The most relevant systems that could benefit from high angular resolution X-ray imaging are the systems that are fairly compact in TeV and spatially coincident with the X-ray PWN (see e.g., PSR J1849-0001 \cite{2024ApJ...968...67G}), leading to a high likelihood that they are associated. Several of these systems show under-luminous X-ray PWN (see e.g., \cite{2013uean.book..359K}) that lack the typical jet-torus morphology seen in other systems. High angular resolution X-ray imaging of the inner regions of these PWN, and those with more typical luminosities and morphologies, could help to elucidate the connection between these systems and their corresponding TeV emission. 

Another more recently discovered class of extreme  TeV emitting objects (i.e., photon energies $E>100$ TeV or electrons with $\sim$ PeV energies) are the BH X-ray binaries with jets \cite{2025NSRev..12af496L}. A recent study has suggested that the radio bright, high duty cycle systems are more likely to be detected at these energies \cite{2026arXiv260706360O}. Similar to PWN the TeV emission appears to be fairly offset from the binaries (see e.g., Table 1 in \cite{2026arXiv260706360O}) . High angular resolution X-ray and radio imaging of the jets at launch and as they escape the binary could help to constrain the particle energy distribution in these jets and how these particles cool over time, which would help place constraints on the particle populations responsible for the TeV emission. Additionally, at late times, the particles in the jet can interact with the ISM leading to a rebrightening in X-rays, which was seen for the TeV detected black-hole binary MAXI J1820+070 \cite{2020ApJ...895L..31E}. High angular resolution X-ray imaging may resolve the shocks in these jets as they interact with the ISM and place constraints on their particle energy distribution, and whether they are re-accelerated in these shocks.

High-resolution X-ray imaging would also be complementary to high-energy neutrino observations. As neutrinos directly trace hadronic interactions, identifying their astrophysical sources is critical for determining where Galactic cosmic rays are accelerated. Current neutrino localizations can encompass multiple potential X-ray and $\gamma$-ray emitters, including pulsar wind nebulae, supernova remnants, and accreting compact objects, which are all candidate neutrino production sites (see e.g., \cite{2017ApJ...836..159D,2007ApJ...665L.131G,2003ApJ...589..481A}). Resolving these environments in X-rays would identify shocks and compact accelerators within the neutrino localization region and determine whether the X-ray morphology and spectrum are consistent with sites of efficient particle acceleration. Coordinated X-ray, $\gamma$-ray, and neutrino observations could therefore distinguish leptonic from hadronic accelerators and identify the Galactic sources capable of producing the highest-energy particles.

\subsection{Winds and Jets} 

\label{sec:wind_jet}

Relativistic jets from X-ray binaries provide nearby laboratories for studying how accretion energy is converted into bulk motion and non-thermal particles. The composition of the jets remains particularly important: a baryon-loaded jet carries substantially more kinetic energy than an electron-positron outflow and can therefore have a much larger impact on its surroundings. SS 433 provides the clearest example of baryonic jets, with Doppler-shifted emission lines from highly ionized species detected in its X-ray spectrum \cite{2002ApJ...564..941M}. Spatially resolved Chandra spectroscopy has further detected Doppler-shifted Fe emission from the arcsecond-scale jets, demonstrating that baryonic material remains hot, or is reheated, far downstream from the binary \cite{2002Sci...297.1673M}. Similar Doppler-shifted X-ray lines have also been associated with relativistic ejecta from the black hole candidate 4U 1630-47 \cite{2013Natur.504..260D}. Milli- to micro-arcsecond imaging combined with spectroscopy would extend these measurements inward toward the launch region, determining where line-emitting material first appears and whether baryons are intrinsic to the jet or are entrained at larger radii. Resolving the X-ray spectrum along the jet would directly map where thermal plasma, shocks, and particle acceleration develop.

Transient ejecta provide a complementary probe of how jets propagate and transfer energy to their environments. Chandra observations of XTE J1550-564 followed relativistic X-ray ejecta over several years and directly measured their deceleration, with synchrotron emission indicating acceleration of electrons to TeV energies in shocks within the ejecta or at their interaction with the surrounding medium \cite{2002Sci...298..196C}. More recently, moving X-ray jets from MAXI J1820+070 showed similar evidence for relativistic propagation, deceleration, and late-time particle acceleration as the ejecta encountered the ISM \cite{2020ApJ...895L..31E}. Higher spatial resolution would allow these ejecta to be detected much closer to the binary and followed from launch through their subsequent expansion and interaction with the environment, linking changes in the inner accretion flow to the eventual deposition of kinetic energy in the ISM. At the same time, highly ionized accretion-disk winds are observed preferentially when relativistic jets are weak or absent, suggesting that winds and jets represent competing or coupled channels through which accretion energy and mass are expelled \cite{2009Natur.458..481N,2012MNRAS.422L..11P}. Spatially resolving these outflows would therefore provide a direct view of how accretion flows divide their energy between winds, relativistic jets, and radiation.

\section{How do Stars Die?}
\label{sec:stellar_death}

\subsection{Supernova Remnants (SNRs)}
\label{sec:star_death_SNR}

\begin{figure}
\centering
\includegraphics[width=0.34\linewidth]{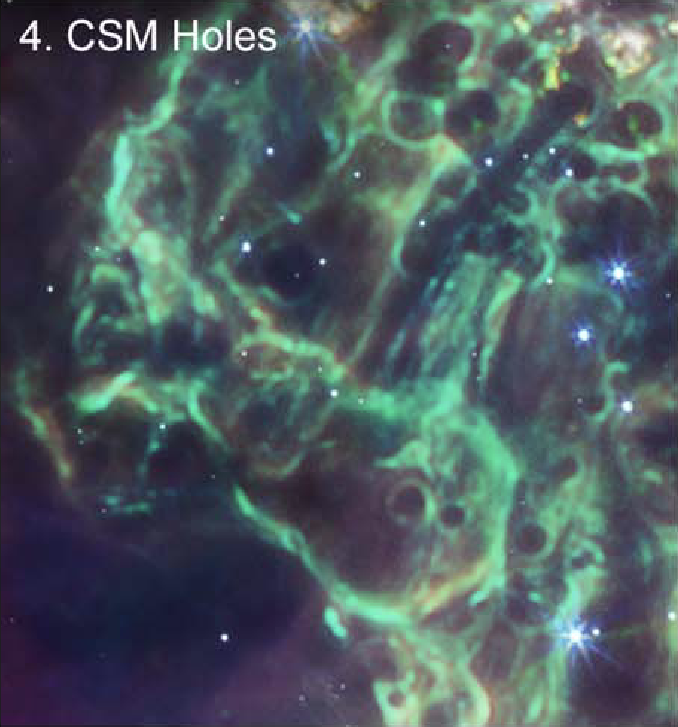}
\includegraphics[width=0.44\linewidth]{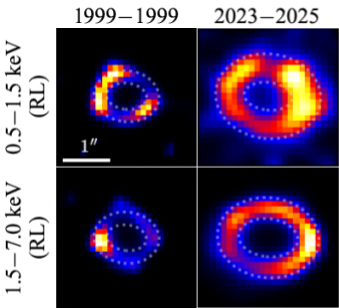}
\caption{ (Left): A JWST combined NIRCam and MIRI image of a region within the so-called ``Green Monster'' of Cassiopeia~A that exhibits $\sim$1''-radius holes created by high-velocity CSM knots, taken from \cite{2024ApJ...965L..27M}. (Right): Deconvolved Chandra soft and hard X-ray images of SN~1987A taken 25 years apart, adapted from \cite{2026ApJ..1006...93S}. }
\label{fig:SNR_structure} 
\end{figure}

SNRs are extremely complex objects: up to tens of parsecs in size and exhibiting structure down to $\lesssim$0.01 pc. For Galactic SNRs, this corresponds to radii of up to tens of arcminutes with sub-arcsecond spatial features.  A JWST survey of Casseiopeia~A revealed structure---including a system of web-like filaments, light echoes, and holes created by high-velocity ejecta knots---down to its MIR diffraction limit of $\sim$0.6'' (0.01~pc; \cite{2024ApJ...965L..27M}; see Figure~\ref{fig:SNR_structure}), and the images hint at structure on even smaller scales. Similarly, Hubble studies of Cas~A identified structures down to the $\sim$0.05'' diffraction limit \cite{2001AJ....122.2644F}. The much lower diffraction limits for X-ray telescopes (0.25 $m$as at 1~keV, 25 $\mu$as at 10~keV) theoretically enable us to obtain images with $\gtrsim$1000x better resolution than optical and IR telescopes. Such observations would reveal information about the supernova's progenitor wind, along with the size and distribution of ejecta knots.

Quasi-stationary flocculi (QSF) in Cas~A have been measured to have radii of 0.004-0.06~pc (0.25''--3.6'' at 3.4~kpc; \cite{2018ApJ...866..139K}). Dust globules in the Crab Nebula have been measured to have radii of $\sim$500~AU (0.15'' at 3.4~kpc \cite{2017A&A...599A.110G}. Observed ejecta knots in Cas~A are 0.2--1.0'' in size \cite{2021ApJ...912...33B}, but \cite{2011ApJ...736..109F} identified flux variability in outer knots with timescales around 1 year or less, indicative of knot scale lengths/sizes of $\lesssim10^{15}$ cm ($\lesssim$0.02''). Overall, 0.1'' spatial resolution can identify most QSFs, dust clumps, and ejecta knots in Galactic SNRs, while a spatial resolution of $\lesssim$0.01'' is required to study them in detail.

Sub-arcsecond spatial resolution would also enable improved studies of SNRs in the Magellanic Clouds, and thus better characterize the progenitors and expansion of SNRs into a lower-metallicity environment. To resolve physical features in LMC \& SMC SNRs at the same level as for Cas~A (D=3.4~kpc) with Chandra, a $\sim$0.03" angular resolution is required, and a field of view (FoV) of few-to-tens of arcminutes$^2$. SN1987A has a diameter of 1.5'': only barely resolved in Chandra \texttt{ACIS} images (see \cite{2016ApJ...829...40F} and Figure~\ref{fig:SNR_structure}). In the upcoming years, the first evidence of reverse shock-heated ejecta and forward shock-heated CSM from this object are expected to be observed. Unfortunately, its small size compared to Chandra's spatial resolution means that spectral signatures from SN1987A overlap; a telescope with higher spatial resolution is vital for spatially isolating different regions of emission in its SN-to-SNR transition period. The other youngest SNR in the Local Group is G1.9$+$0.3, with age estimates of 100--150~yr \cite{2011ApJ...737L..22C,2008ApJ...680L..41R}. This leaves a gap of $\sim$80 years with no observations capturing the early stages of SNR evolution, highlighting the importance of observing SN~1987A over the next few decades. The 0.01~pc-sized structures observed in Galactic SNRs correspond to 0.04'' at 50~kpc, but as the physical features in SN1987a are likely an order of magnitude smaller due to its young age, an X-ray imager with $\sim$1--10~$m$as resolution is required to isolate and study specific features in SN~1987A. For sufficiently nearby galaxies, the same capability would extend beyond supernova remnants to newly discovered supernovae themselves, allowing spatially resolved X-ray observations of the evolving shock, ejecta, and circumstellar interaction on physical scales inaccessible with current X-ray observatories.

Type Ia supernovae provide another opportunity to connect compact-binary evolution directly to the physics of stellar explosions. The progenitor channels of SN Ia remain uncertain, with mergers of double white-dwarf binaries among the leading possibilities. LISA will identify large numbers of compact double-WD systems and should provide a nearly complete census of the shortest-period candidate SN Ia progenitors, using their gravitational-wave chirp masses to constrain the component masses and merger rates \cite{2019MNRAS.482.3656R,2024A&A...691A..44K}. High-resolution X-ray observations would provide a complementary view after explosion, resolving the distribution of shocked ejecta and circumstellar material and searching for asymmetries or structures that distinguish different progenitor and explosion scenarios. Together, gravitational-wave measurements of the pre-explosion binary population and spatially resolved X-ray observations of their remnants would directly connect progenitor demographics to the outcomes of thermonuclear explosions.

Finally, many SNR distances are uncertain: parallaxes to embedded NSs or ejecta knots could help constrain their physical sizes, explosion energies, and expansion rates. A telescope that can localize the positions of point sources to $\sim$0.1~$m$as would enable the parallax measurements of Galactic SNRs.

\subsection{High-Mass Gamma-ray Binaries} 
\label{sec:HMGBs}

High-mass gamma-ray binaries provide a rare view of young compact objects that remain bound to massive stellar companions, offering a direct link between the end stages of massive-star evolution and the formation of neutron stars and black holes. These systems consist of young Be or O-type stars with a compact object in an often highly eccentric orbit (see e.g., \cite{2013A&ARv..21...64D} for a review). In the few systems where the compact object is known to be a neutron star, through the detection of radio pulsations, its high spin-down power and young age imply a strong relativistic pulsar wind. This wind interacts and shocks with the stellar wind produced by the massive companion, accelerating particles that produce synchrotron X-ray emission and TeV emission through inverse-Compton scattering of photons from the companion star. The canonical system is PSR B1259-63 \cite{1992ApJ...387L..37J}, where a young energetic pulsar is in a 3.4-year orbit around a massive Be star and crosses the companion's decretion disk twice per orbit \cite{2005MNRAS.358.1069J,2006MNRAS.367.1201C}. At its distance of 2.6~kpc, the projected orbit has a size of about 1~mas \cite{2018MNRAS.479.4849M}, so an X-ray observatory with sub-mas angular resolution could directly resolve the X-ray emission site as the pulsar moves through its orbit. Because these systems are young, measurements of the nature of the compact-object, orbital eccentricity, three-dimensional motion, and interaction geometry can also preserve information about the mass loss and natal kick associated with the supernova that formed the compact remnant, providing constraints on how massive stars end their lives. This would be particularly useful for systems where no radio pulsations have been identified.

\begin{figure}[ht]
    \centering
    \includegraphics[width=0.73\linewidth]{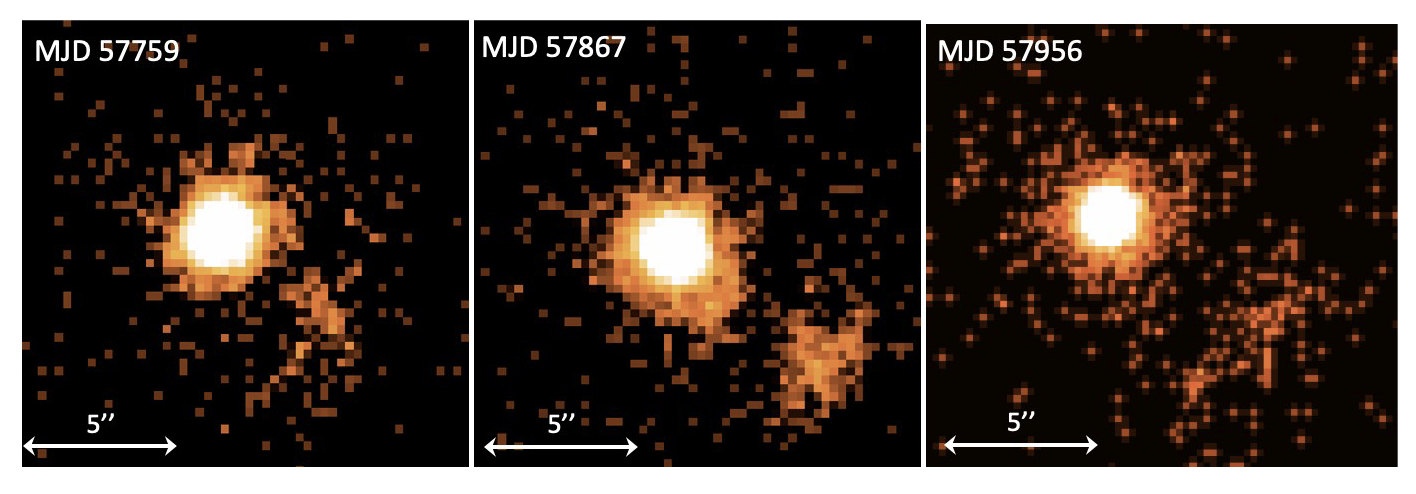}
    \includegraphics[width=0.4\linewidth]{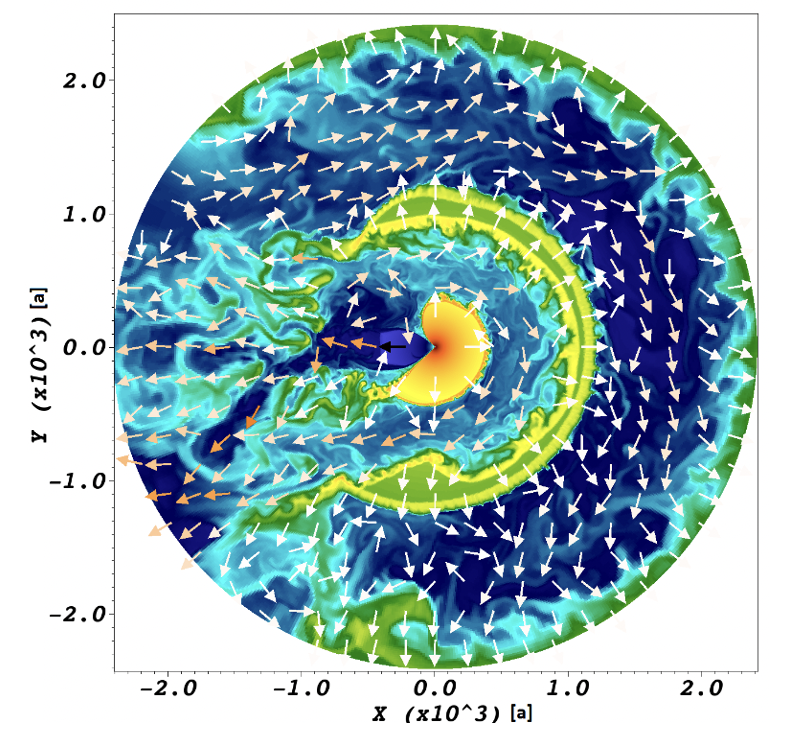}
    \caption{Top: Extended X-ray emission launched from the PSR B1259-63 binary as seen by Chandra. Figure adopted from \cite{2019ApJ...882...74H}. Bottom: Hydrodynamical simulation of the interacting pulsar and stellar winds in the PSR B1259-63 system. The X and Y axes are shown in units of the orbital semi-major axis ($a\sin{i}\approx1296$ lt-s; \cite{2018MNRAS.479.4849M}) corresponding to angular scales of several hundred $m$as  at PSR B1259's distance. The colors show the density distribution of the winds, with the tenuous pulsar wind shown in blue and the dense stellar wind shown in yellow/green. The arrows show the direction of the wind flow with darker colors representing a faster flow speed. Figure adopted from \cite{2016MNRAS.456L..64B}.}
    \label{fig:b1259_image}
\end{figure}

One of the most interesting features exhibited by these sources is extended X-ray emission near the binary (see e.g., \cite{2011ApJ...735...58D,2011ApJ...730....2P}). This was originally thought to be from jets (assuming a black hole scenario) or the termination shock of a PWN. However, follow-up studies of the extended emission surrounding PSR B1259-63 showed that material was being expelled from the binary at a high velocity ($\sim$0.1c), with possible hints of acceleration \cite{2015ApJ...806..192P,2019ApJ...882...74H}. The predominant theory for this emission is that, due to the high eccentricity of the orbit, the pulsar spends a majority of its time near apastron. The stellar wind is dynamically dominant over the pulsar wind, but the relativistic pulsar wind carves out a channel in this dense, slow moving stellar wind. As the pulsar passes through the disk it knocks disk material into this channel, which is shocked by the pulsar wind (producing synchrotron emission observable at X-ray energies) and possibly accelerated, eventually being ejected from the binary \cite{2015ApJ...806..192P,2016MNRAS.456L..64B,2019ApJ...882...74H}. Hence, the extended X-ray emission is observed to move along the apastron direction (see Figure \ref{fig:b1259_image}). To date, this phenomena has been observed after 2 of the 4 periastron passages that were closely monitored by Chandra \cite{2023ApJ...958....5H} and it is still unclear what exactly leads to the appearance/disappearance of this emission. High-angular resolution imaging of this binary during (and following) perisatron passage could help us better understand the launching mechanism of this disk material and what dictates whether the material will be launched after a given periastron passage. Hydrodynamical simulations suggest that the shocked wind structures close to the NS and its companion should be resolvable at distances 100s of times the orbital semi-major axis (or several hundred $m$as) a few hundred days after periastron passage (see Figure \ref{fig:b1259_image}). Additional observations of other systems where extended emission has been observed, such as LS 5039 \cite{2011ApJ...735...58D}, could help to determine whether it has a similar origin or is due to an entirely different process (e.g., jets). 

\subsection{Natal Kicks}
\label{sec:natal_kicks}

Neutron stars are observed to have average sky motions of a few hundred km s$^{-1}$ \cite{2005MNRAS.360..974H,2025ApJ...985...12W,2025A&A...700A..75D}, larger than can be explained from the disruption of a progenitor binary. Understanding the link between NS kick velocities and SNR properties -- asymmetry, energy, magnetic field, progenitor mass, etc. -- can help elucidate the processes that occur within and lead to supernova explosions. For example, \cite{2017ApJ...844...84H,2018ApJ...856...18K} showed that NSs preferentially move in a direction opposite the bulk motion of ejecta, supporting a conservation of momentum-like argument for the acceleration of NSs. Simulations by e.g., \cite{2024ApJ...963...63B} have supported these observational findings, but additionally indicated that the slowest moving NS from the smallest progenitors (v$\sim$100~km s$^{-1}$; M$\lesssim$10M$_\odot$) might be accelerated as a result of asymmetries in bulk neutrino emission. 

\begin{figure}[h]
\centering
\includegraphics[width=0.45\linewidth]{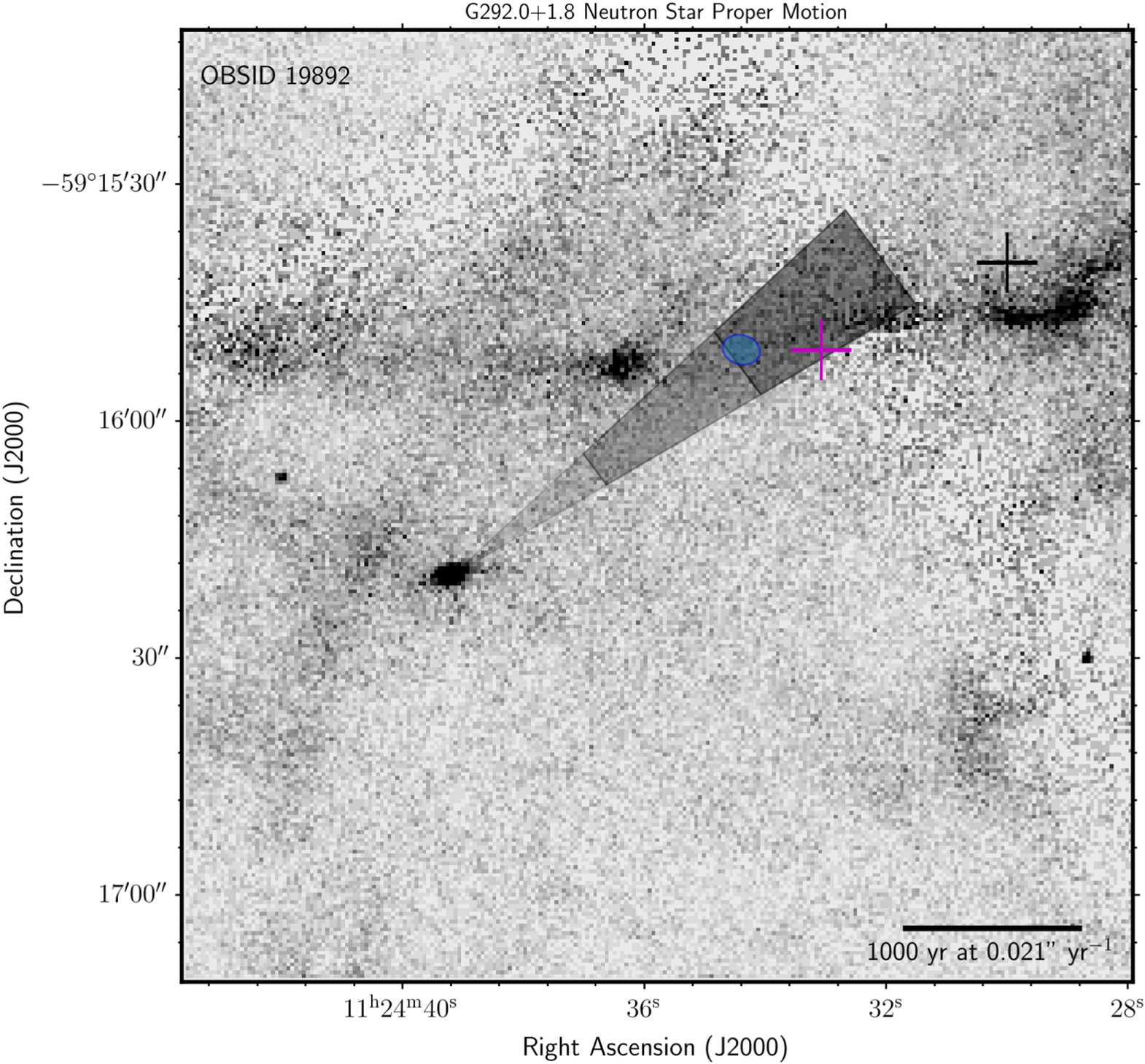}
\includegraphics[width=0.54\linewidth]{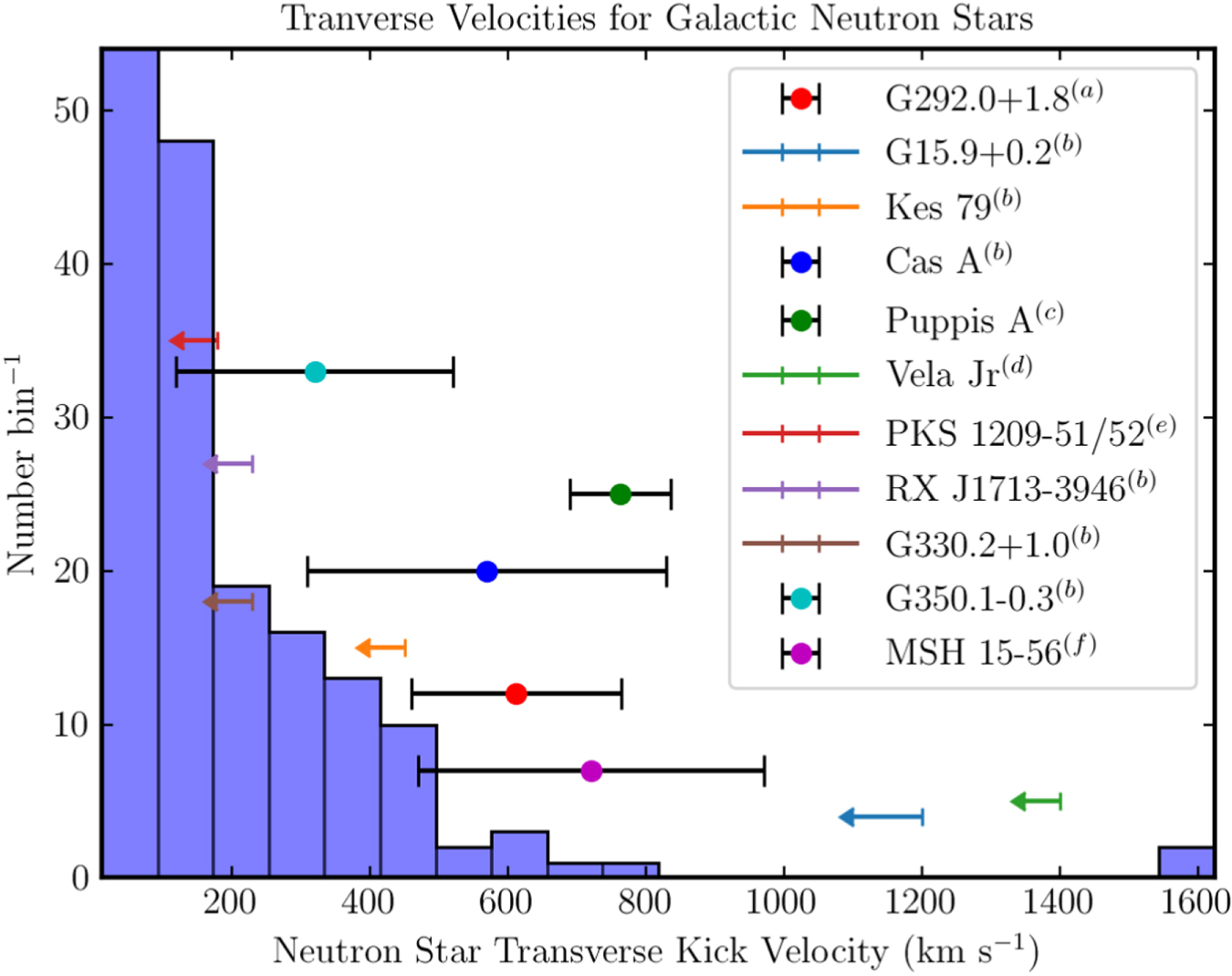}
\caption{ (Left): The backward evolution of the NS in the SNR G292.1$+$1.8. Each shaded region represents 1000 years, the blue ellipse is the SNR's center of expansion, the magenta plus is the SNR's geometric center, and the black plus is the center of X-ray emission. (Right): Distribution of velocities for the bulk pulsar population compared to recent kick velocity measurements of individual NSs associated with SNRs. Figures are taken from \cite{2022ApJ...932..117L}.}
\label{fig:NS_Vels} 
\end{figure}

A NS velocity of $\sim$400 km s$^{-1}$ corresponds to a sky motion of $\sim$0.03" yr$^{-1}$ at a distance of 3~kpc. The diagonal lines in Figure \ref{fig:resolution_limits} show the angular displacement expected for representative transverse velocities over fixed observing baselines, illustrating the range of natal-kick motions that could be measured as a function of source distance. Currently, obtaining accurate measurements of Galactic NSs' motions requires Chandra baselines of $\gtrsim$10~years and the ability to detect a sufficient number of registration sources to perform absolute astrometric corrections, and often still results in $>$100~km/s uncertainties (e.g., \cite{2021A&A...651A..40M}). An example of this approach is shown in the left panel of Figure \ref{fig:NS_Vels} for G292.1+1.8, where a proper motion corresponding to a velocity of 612 $\pm$ 152 km s$^{-1}$ was measured using a Chandra baseline of 10 years. The authors compared the neutron star's backward motion to independently determined centers of the remnant to obtain a better constraint on the SNR's age \cite{2022ApJ...932..117L}. The difficulty in obtaining robust NS velocity measurements has limited the ability to build large samples for study. To date, there are only $\sim20$ robust velocity measurements for NSs associated with SNRs, several of which are shown in the right panel of Figure \ref{fig:NS_Vels}, and only about half of these remnants are dominated by thermal ejecta emission rather than swept-up material or nonthermal synchrotron emission from the forward shock and/or PWN. These are a small fraction of the hundreds of known Galactic SNRs \cite{2025JApA...46...14G}, and are heavily biased to nearby and bright sources.

Identifying NS positions to within 1 $m$as would enable the proper motion measurement of all NSs on this side of the Galactic center (d$\lesssim$8~kpc) to within $\pm$50 km s$^{-1}$ with just 1-year baselines. To make such a measurement within a single observing cycle (e.g., a baseline of 3 months), centroid uncertainties closer to 0.25 $m$as are necessary. Furthermore, as the positions of point sources can be identified to greater precision than the pixel size and PSF of a detector (roughly $\sim$10x, assuming a sufficient pointing accuracy and number of counts), a detector would only need to have $\sim$2--10 $m$as spatial resolution to make these measurements.

A future high-resolution X-ray observatory would also benefit from the long time baseline provided by the existing Chandra archive. For a mission launched around 2040, many neutron stars observed during the first decade of Chandra operations would have astrometric baselines of $\sim30$--40 years. Over such intervals, even relatively modest improvements in absolute astrometric precision could translate into useful proper-motion constraints. For sources within $\sim8$~kpc, combining a future position measurement with a Chandra position registered to the Gaia reference frame could constrain transverse velocities to $\lesssim100$~km/s for systems with first-epoch positional uncertainties of order $0.1''$. This would substantially enlarge the sample of neutron stars with measured transverse velocities without requiring repeated observations over the lifetime of the future mission.

For neutron stars in the Magellanic Clouds, their $\sim50$~kpc distances correspond to typical transverse motions of only $\sim17$ $m$as per decade, making direct X-ray proper-motion measurements extremely challenging with Chandra. In some systems, however, complementary optical astrometry may be possible. Optically detected young pulsars can in principle be measured directly, while high-mass X-ray binaries generally have bright stellar companions whose motions may be accessible with Gaia. High-precision X-ray astrometry would nevertheless provide a complementary and in some cases uniquely direct measurement of the compact object's motion, particularly for isolated or optically faint neutron stars and for systems in which the stellar companion does not faithfully trace the natal motion of the remnant. Assuming the ability to identify the centroid of a point source to a factor of $\sim$10x better than the spatial resolution/pixel size, a telescope with spatial resolution of 0.4~mas would unlock 1$\sigma$ uncertainties of $\pm$50 km s$^{-1}$ for all Magellanic Cloud NS velocities within a single observing cycle (e.g., a 3-month baseline), while a telescope with a spatial resolution of 6~mas would enable us to constrain all Magellanic Cloud NS velocities to $\pm$50 km s$^{-1}$ using a reasonable mission-length baseline of 4~years.

Kick velocity distributions of black hole X-ray binaries remain more controversial. This is partly a result of their smaller numbers and partly because they tend to reside in systems with low-mass companions that are fainter in quiescence \cite{Casares2014, CorralSantana2016, Tetarenko2016, Gandhi2019, Fortin2024}. Pinning down black hole kicks is arguably even more important than for neutron stars, because the canonical assumption is that kicks will be moderated by `fallback' of material on to a proto-neutron star core during supernova, resulting in weak momentum recoil which will scale inversely with compact object mass \cite{Fryer2012}. By contrast, kinematic studies using radio interferometric observations and optical astrometry are suggesting that at least some accreting black hole systems require strong kicks \cite{Mirabel2002, Willems2005, Gandhi2019, 2019MNRAS.489.3116A, Kimball2023, dashwoodbrown24}, but not necessarily all \cite{Burdge2024, Nagarajan2025}, with potential inferences for a bimodal distribution of kicks dependent on the mass of the binary \cite{zhao23}. Identifying more systems and tracing their proper motions is key to pinning down the kick distribution. X-rays can help with both discovering more accreting systems, and also with tracing their kinematics in obscured regions of the Galactic disc/bulge beyond the reach of current optical surveys.

\subsection{Black Hole and Neutron Star Orbits: Masses, Spins and Misalignments} 
\label{sec:BH_NS_orbit}

Understanding the masses of black holes and neutron stars in X-ray binaries is crucial for understanding the supernovae that formed them.  In most cases, obtaining a $v\;{\rm sin}\; \theta$ measurement is straightforward.  The harder ingredients, the ones that break degeneracies in the mass function and unlock individual mass measurements, are the inclination angles and mass ratios of the binaries.  While these can often be estimated through ellipsoidal modulations and rotational broadening, there are systematic uncertainties for both.  Furthermore, neither works in a straightforward manner for wind-fed detached binaries.  Astrometric wobble can provide both of these values.  Black hole masses are important for determining whether the black holes formed from prompt collapses of stars or a supernova with fallback accretion, while precise neutron star masses can constrain the difference between formation via iron core collapse and electron capture \cite{2004ApJ...612.1044P, 2017ApJ...846..170T}.

Additionally, the disk continuum method for estimating black hole spins relies on having accurate values of the distance, inclination angle, and black hole mass for a system \cite{2014SSRv..183..295M, 2014ApJ...790...29G}.  The reflection method can provide black hole spin values without any additional system knowledge \cite{2014SSRv..183..277R, 2023ApJ...946...19D}, but independent measurements of the inclination angle provide a vital check on the model systematics of reflection fitting. Finally, measurement of the position angle and inclination angle of the binary orbit relative to the inclination angle and position angle of relativistic jets can be used to determine whether the jet launching direction is perpendicular to the orbital plane; if it is stably in some other direction, that is presumably because it goes along the black hole spin axis, which is misaligned from the orbit due to frame dragging.  Hints of this happening come from inclination angle offsets \cite{2001ApJ...553..955F,2002MNRAS.336.1371M} and from polarization measurements\cite{2022Sci...375..874P}, as well as from the interpretation of certain quasi-periodic oscillations as Lense-Thirring precession \cite{2001ApJ...553..955F}.

For a compact object of mass, $M_{CO}$ in a binary with a star of mass $M_*$, for a total mass of $M$, at period $P$ and distance $d$.  The angular size of the semimajor axis of the orbit is:
\begin{equation}
    \Delta \theta = 1~m{\rm as} \left(\frac{P}{\rm yr}\right)^{2/3} \left(M/M_\odot\right)^{-2/3} (M_*/M_\odot) \left(\frac{d}{\rm kpc}\right)^{-1}.
\end{equation}

For a 10 $M_\odot$ black hole, with a 1 $M_\odot$ companion at 5 kpc in a 1 day orbit, this will be just under a $\mu$as.  For neutron stars in wide orbits, it can be considerably larger.  Because these are very bright sources, a signal to noise ratio of about 1000 may be obtained in many cases, so sufficient centroiding accuracy to measure the orbit is likely to be possible with resolution of the order $\mu$as for many objects.

\section{How is Accretion Fueled?}  
\label{sec:accretion_fuel}

\subsection{Magnetically Channeled Accretion} 
While T-Tauri systems facilitate the investigation of magnetic channeling at the stellar
scale ($\sim7\times10^8$ m), and white dwarf systems ($\sim6\times10^6$ m) enable studies at planetary scales (e.g., see \cite{hires_stellar}), the
analysis of magnetically channeled accretion onto neutron stars offers a unique laboratory
where strong-field General Relativity (GR) dominates. The study of X-ray emission from
this environment can provide critical constraints on the fundamental Equation of State
(EoS) of dense nuclear matter and the most intense magnetic fields observed in the
Universe. Advancing the understanding of these processes requires a significant leap
in observational capabilities to resolve emission at the neutron star scale ($\sim 10^4$ m),
corresponding to an angular resolution of $\lesssim10^{-10}$ arcseconds for a distance of 1 kpc. 

The study of neutron star accretion is particularly compelling for systems
characterized by magnetic field strengths in the $10^{12} - 10^{15}$ G regime. In this domain, the
emission physics is highly sensitive to the accretion rate; the emission geometry is
predicted to transition from a pencil beam (originating from the accretion shock at the
stellar surface) to a fan beam (originating from the lateral walls of the accretion
column) as the accretion rate and density increase. The resulting X-ray emission from the
accretion column may extend vertically to tens of kilometers above the neutron star
surface \cite{Basko76,BeckerWolff07}. At moderate accretion rates, the intense magnetic
field truncates the accretion disk at the Alfven radius, resulting in prominent X-ray
emission from the inner disk on scales of 0.1$-$0.01 $\mu$as
\cite{2010_a0535_fek,2026_herx1_fek}. In principle, spatially resolving X-ray emission from the neutron-star surface could reveal both magnetic poles through gravitational light bending and provide a direct probe of strong-field general relativity.  Nearby bright millisecond pulsars, with existing emission topology constraints from NICER, would provide the most favorable systems for such measurements \cite{psrj0437_nicer}. %At a distance of 1~kpc, a neutron star with a characteristic diameter of $\sim20$~km subtends $\sim1.3\times10^{-10}$~arcsec, while a $\sim1$~km hot spot subtends $\sim6.7\times10^{-12}$~arcsec.

Furthermore, the study of cyclotron absorption
features at higher energies allows for direct empirical constraints on the magnetic field
for neutron stars with B $\sim 10^{12}$  G. Notably, there are accreting neutron stars exhibiting
candidate proton cyclotron lines within the soft X-ray band, potentially facilitating the study of magnetic fields in excess of the Schwinger limit \cite{2019staubert_cyc}. It has also
been established that the central engines of many Ultra Luminous X-ray sources (ULXs) are
neutron stars \cite{Bachetti14,furst16,israel17}. These objects accrete at super-Eddington
rates, conditions under which the accretion flow geometry is expected to undergo
fundamental structural changes (e.g., \cite{2017_mushtukov}).  Transient examples of super
Eddington accretion onto neutron stars are known within the Galaxy
\cite{2019_tao_swj0243,2022_liu_swj0243}.

\subsection{Accretion Feeding}  

\label{sec:accretion_feed}

\begin{figure}[ht]
\centering
\includegraphics[width=0.53\linewidth]{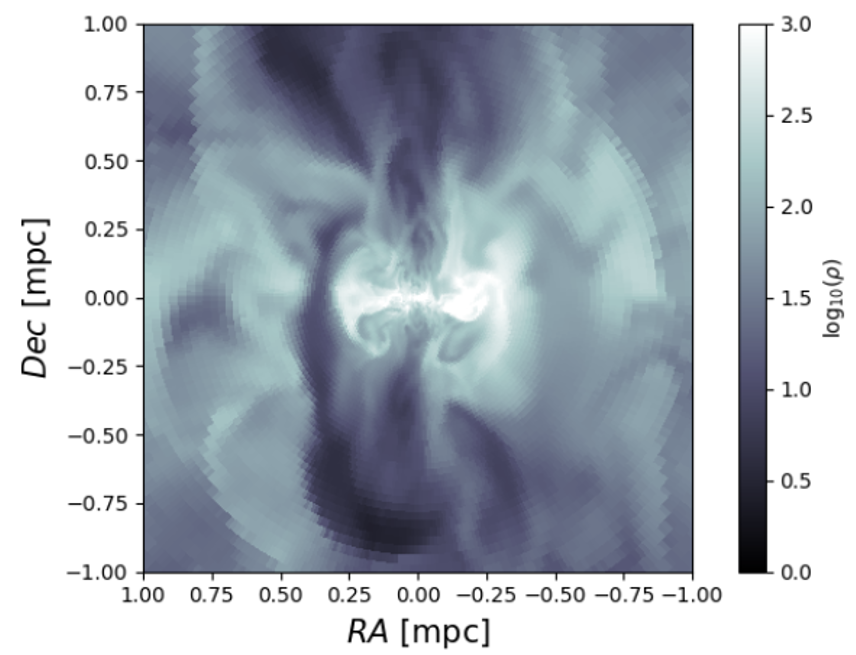}
\caption{Predicted morphology of the wind-fed accretion flow around \SgrA. The panel shows a plane-of-sky mass-density slice from a MHD simulation and spans $\pm 1$ mpc, corresponding to approximately $\pm 25$ $m$as or $\pm 5\times10^{3} R_g$. The simulation produces rotating, asymmetric flows with lower-density cavities whose orientation and structure depend on the magnetic field and stochastic wind feeding. Milli-arcsecond X-ray imaging would resolve this region and test these predicted density structures, distinguishing quasi-spherical inflow from rotationally and magnetically shaped accretion and outflow. Adapted from Figure 1 of \cite{Ressler2023}.}
\label{fig:sgra_accflow_sim}
\end{figure}

Sgr~A$^*$ is extraordinarily faint, placing it in the category of low-luminosity active galactic nuclei (LLAGN), a regime that likely represents the most common state of supermassive black-hole accretion in the local Universe \cite{Ho2008}. 1 $m$as X-ray imaging would bridge the current gap between the resolved hot-gas reservoir at $R_{\rm Bondi} \sim 10^5-10^6\ R_g$ (resolvable with Chandra) and the unresolved inner accretion flow down to $200\ R_g$, and help decipher why material does not reach the inner regions. Within the Bondi radius ($R_{\rm Bondi} \sim 4$''), hot plasma supplied by stellar winds is gravitationally captured, but the connection between this large-scale reservoir of hot gas and the inner accretion flow remains unconstrained. Despite $10^{-5}\ \rm M_\odot yr^{-1}$ of available material at the Bondi radius \cite{Cuadra2008}, Sgr~A$^*$'s accretion rate close to the event horizon is only $10^{-8}\ \rm M_\odot yr^{-1}$ \cite{Ressler2020}, with models showing that $< 1\%$ of inflowing gas reaches the SMBH \cite{Yuan2014}. Hydrodynamic simulations find that around 0.2'', or $4\times 10^4\ R_g$, the flow transitions from quasi-spherical to a rotating outflow-dominated structure. Milli-arcsecond X-ray imaging would provide the first spatially resolved measurement of this transition between $10^{2}$--$10^{4}\,R_g$, where simulations predict asymmetric, rotating flows and lower-density polar cavities (Figure~\ref{fig:sgra_accflow_sim}). X-ray observations currently indicate that the flow is strongly mass-losing at large radii; spectral modeling of the quiescent emission suggests an inflow-outflow balance near $R_{\rm Bondi}$ \cite{Balakrishnan2024, Wang2013}. But closer in, less material is proportionally ejected \cite{Ma2019}. Obtaining a constraint of the dependence of density and temperature on radius would provide insight into quasi-spherical flows and other LLAGN. This science goal requires a FOV of 4'' $\times$ 4'' and an energy resolution of 100 eV at 6 keV at a minimum, but if possible an energy resolution of $\sim$ 10 eV (about 500 km/s precision) would separate fine structure seen in the Chandra spectrum ($\sim$ 150 eV at 6 keV) and allow us to compare to hydrodynamical simulations \cite{Ressler2018}. An effective area of $\sim$ 10$^4$ cm$^2$ is required to conduct a spatially resolved spectroscopic analysis.

\subsection{Compact Binary Feeding}
\label{sec:comp_bin_feed}

In interacting binaries, the geometry of mass transfer determines how material reaches the compact object, how angular momentum is redistributed, and whether matter is ultimately accreted or expelled from the system. In Roche-lobe overflow (RLOF) systems, the mass-transfer stream, outer accretion disk, and possible circumbinary outflows all occupy scales comparable to the binary separation, although the bulk of the X-ray luminosity in ordinary states is expected to originate much closer to the compact object. The most promising opportunities for spatially resolving larger-scale X-ray structure may therefore occur during outbursts, when the accretion flow becomes highly dynamic and strong disk winds, scattering material, stream-disk interactions, and ejecta can redistribute X-ray emission over a much broader range of radii. V404~Cyg provides a useful fiducial example: for $d=2.39$~kpc, $P\simeq6.5$~days, $M_{\rm BH}\simeq12\,M_\odot$, and $M_2\simeq0.7\,M_\odot$, the binary separation is $\simeq0.16$~AU, corresponding to $\sim70~\mu$as. An angular resolution of $10~\mu$as would probe $\sim0.024$~AU, providing multiple resolution elements across the binary. During bright outbursts, such observations could search for asymmetric or extended X-ray emission associated with disk winds, the stream-impact region, and material expelled from the binary, while testing whether non-conservative RLOF produces larger-scale circumbinary structures \cite{2025ApJ...990..172S, 2025arXiv251024127S}.

A complementary regime is provided by detached and semi-detached systems in which the compact object accretes from the wind of its companion. In high-mass X-ray binaries, Cygnus~X-1 is a key fiducial case \cite{2021Sci...371.1046M}: its $\sim0.4$~AU binary separation corresponds to $\sim200~\mu$as at 2.2~kpc, while $10~\mu$as imaging probes scales of only $\sim0.02$~AU. For comparison, one gravitational radius for the $\sim21\,M_\odot$ black hole in Cygnus X-1 subtends $\sim10^{-10}$~arcseconds. Such observations could spatially connect the black hole environment to the focused stellar wind, testing how an asymmetric wind develops into an accretion stream or disk and how changes in the donor wind propagate inward to modulate the X-ray luminosity. Similar physics operates in symbiotic binaries, where a white dwarf or neutron star accretes from the dense, slow wind of an evolved giant. Three-dimensional hydrodynamical modeling-calculations show that gravitational focusing can produce complex streams, spiral-like structures, and accretion rates substantially larger than predicted by simple Bondi--Hoyle--Lyttleton capture \cite{2017MNRAS.468.3408D}. For the nearest and widest interacting binaries, $m$as to sub-$m$as X-ray imaging could therefore resolve the structures through which wind material is captured and circularized, as well as follow changes in those structures during accretion-driven outbursts and jet ejections.

These measurements have implications extending beyond the instantaneous accretion process. Interacting binaries containing accreting white dwarfs are possible progenitors of planetary nebulae and, when the white dwarf can retain sufficient accreted mass, potential contributors to the single-degenerate channel for Type~Ia supernovae. Directly measuring how much transferred material is accreted, expelled in winds, or carried away by episodic jets is therefore central to determining whether massive white dwarfs can grow toward the Chandrasekhar mass and to understanding how binary interaction shapes the circumstellar environment prior to stellar death. More broadly, resolving the formation of disks, focused streams, winds, and jets in nearby compact binaries would establish spatially resolved laboratories for accretion processes that operate across a much wider range of mass scales, from young stellar objects to active galactic nuclei.

\section{What are the Populations of X-ray Sources?} 
\label{sec:xray_pop}

Ultra-high-resolution X-ray imaging would enable population studies across the full range of Galactic environments, from the nuclear star cluster to globular clusters and the Galactic disk, determining how the relative numbers of cataclysmic variables, white dwarf binaries, ultracompact X-ray binaries, neutron star systems, and black hole binaries depend on stellar density, age, metallicity, and dynamical interactions.

\subsection{Galactic Center}
\label{sec:galactic_center}
\begin{figure}
    \centering
    \includegraphics[width=0.38\linewidth]{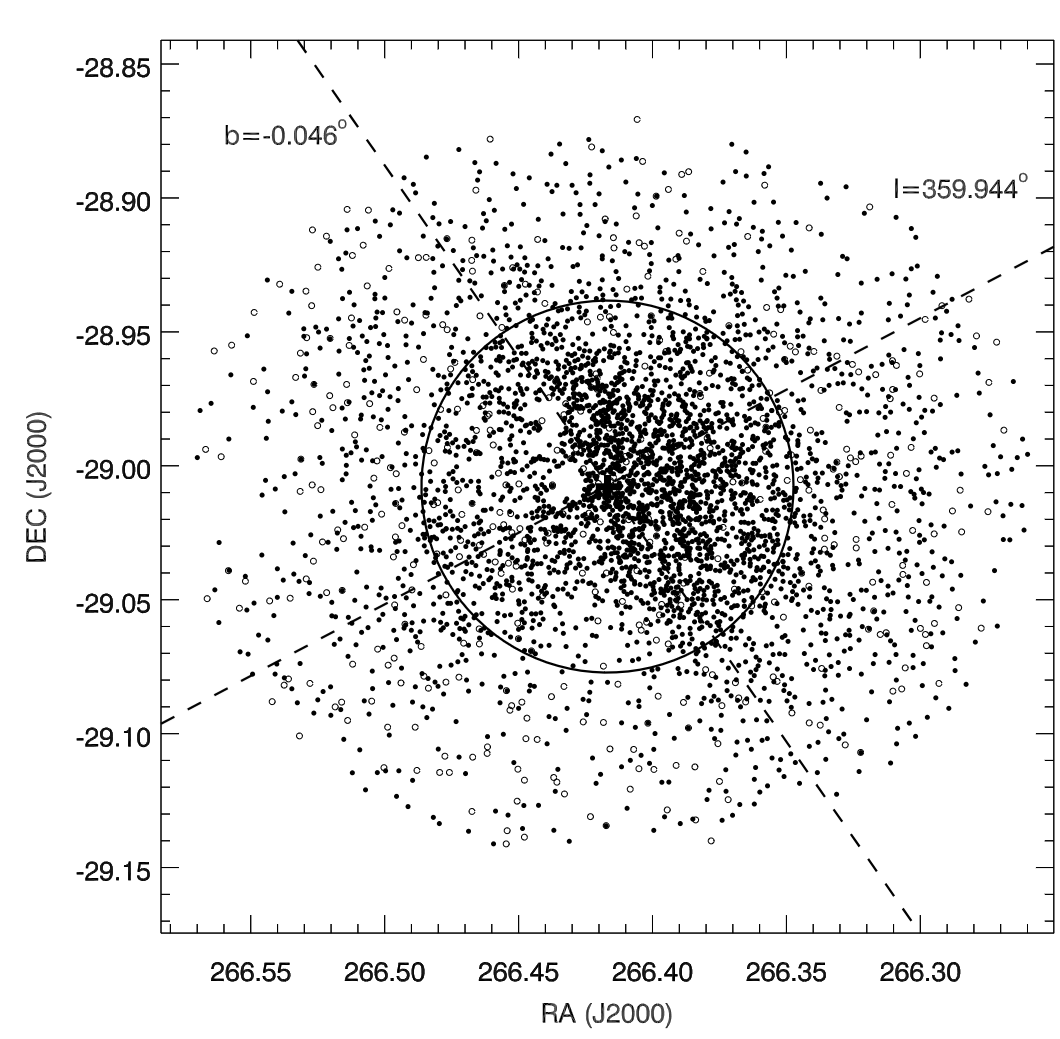}
    \includegraphics[width=0.48\linewidth]{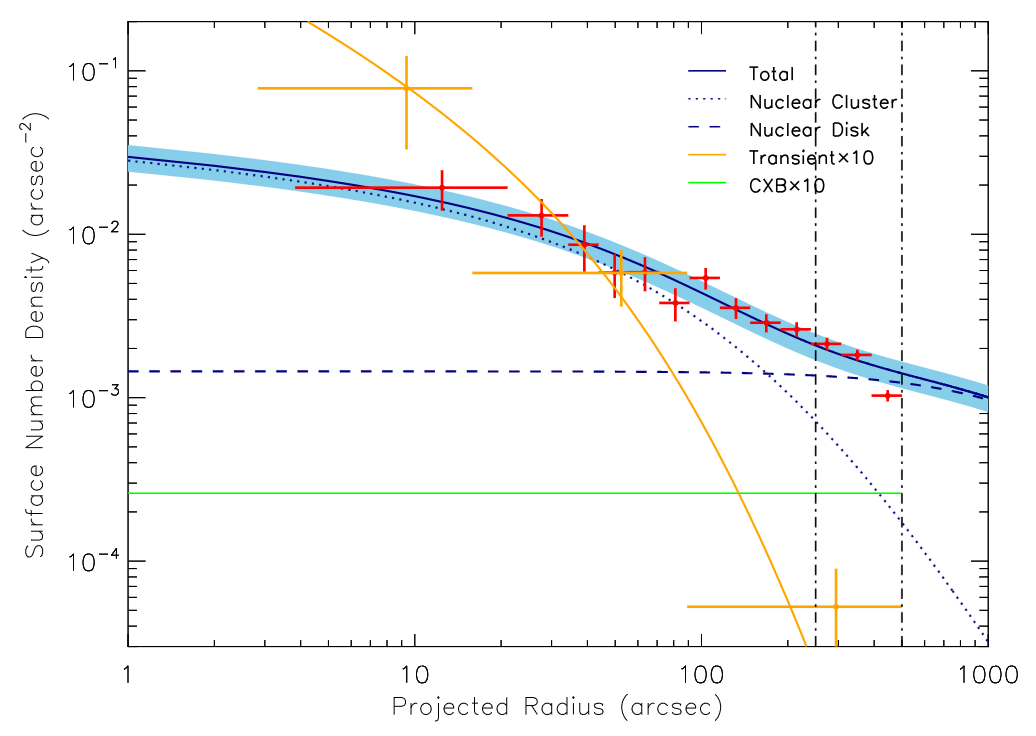}
    \caption{Chandra measurements of the Galactic Center X-ray-source population. (Left): Spatial distribution of the 3619 sources detected within $500''$ of Sgr~A$^*$. Filled points denote sources located at the Galactic Center or beyond, while open circles denote likely foreground sources; the large circle marks a projected radius of $250''$. (Right): radial surface-density profile of a flux-limited sample compared with models for the nuclear star cluster and nuclear stellar disk. The known X-ray transients, plotted with their density multiplied by ten for clarity, exhibit a substantially steeper central distribution than the overall source population. These measurements illustrate both the extreme projected source density and the possible radial variation in source demographics that motivate higher-resolution, class-resolved studies. Adapted from Figures 3 and 12 of \cite{Zhu2018}.}
    \label{fig:gc_sources}
\end{figure}

The Galactic Center is home to thousands of X-ray point sources \cite{2009ApJS..181..110M,2003ApJ...589..225M} and provides a unique laboratory for determining how compact-object populations depend on stellar density, age, metallicity, and dynamical interactions. Deep Chandra observations have detected $\sim$2000 X-ray sources within the central $17' \times 17'$ \cite{2003ApJ...589..225M}, while a wider $2^\circ \times 0.8^\circ $ survey detected $\sim$9000 sources \cite{2009ApJS..181..110M}. An ultradeep survey of the central 500'' subsequently detected 3619 sources (Figure~\ref{fig:gc_sources}, left), approximately 3500 of which are probable GC sources, down to an intrinsic $L_{\rm 2-10\ keV} \sim 10^{31}$ erg s$^{-1}$\cite{Zhu2018}. These sources include magnetic and non-magnetic cataclysmic variables (CVs), quiescent and transient neutron stars and black hole X-ray binaries, pulsars and magnetars, massive stars and colliding-wind binaries, and contaminating sources (foreground stars or background AGN). Distinguishing magnetic and non-magnetic CVs from faint neutron-star binaries and UCXBs would establish whether the nuclear source population is primarily an extension of the Galactic-bulge population or is instead strongly modified by dynamical interactions near Sgr~A$^*$.

Despite extensive surveys, the nature of most of the sources remains unknown. Many sources contain too few photons for detailed spectroscopy, and different source classes can produce similarly hard and highly absorbed X-ray spectra. Hard X-ray observations suggest that magnetic CVs account for 40-60\% of the brighter NuSTAR population \cite{Hong2016}, but the relative contributions of non-magnetic CVs, neutron star or black hole binaries, pulsars, and massive stars is poorly constrained. Multiwavelength follow-up provides critical additional information. For example, infrared spectroscopy confirmed 16 massive stellar counterparts to Chandra sources, including Wolf-Rayet stars and supergiants \cite{Mauerhan2010}. Nevertheless, the high density of infrared sources means that a Chandra positional region can constrain multiple possible counterparts, preventing secure classifications for much of the population.

The source population may also change substantially with distance from Sgr~A$^*$ (Figure~\ref{fig:gc_sources}, right). Four of the seven X-ray transients detected within 23 pc of Sgr~A$^*$ were located within the central parsec, corresponding to a strong excess per unit stellar mass relative to the surrounding region \cite{Muno2005}. A concentration of 12 quiescent X-ray binary candidates in the same region has been interpreted as evidence for a cusp of stellar-mass black-hole binaries \cite{Hailey2018}. By contrast, the wider inner-$20$ pc population appears to be dominated by CVs, with quiescent low-mass X-ray binaries contributing at most a few percent \cite{Zhu2018}.
Milli-arcsecond X-ray imaging could securely associate an X-ray source with an individual infrared or radio counterpart and separate faint sources from nearby objects, diffuse plasma, and contamination from the PSF wings of bright transients. With stable sub-$m$as relative astrometry, this resolution could also enable proper-motion measurements for sufficiently bright X-ray sources. A transverse velocity of $100$ km s$^{-1}$ in the nuclear star cluster corresponds to approximately $2.6$ $m$as yr$^{-1}$ at 8.1 kpc. Multi-epoch X-ray astrometry could therefore help distinguish Galactic Center members from foreground and background sources and determine whether white dwarfs, neutron stars, and black-hole binaries have different radial and kinematic distributions around Sgr~A$^*$. These measurements would test mass segregation, dynamical binary formation around a supermassive black hole, compact-remnant retention, and predictions for gravitational-wave progenitors in galactic nuclei. A class-resolved luminosity function would also determine which populations produce the unresolved hard X-ray emission from the nuclear star cluster.

A particularly compelling application of ultra-high-resolution X-ray imaging would be to extend the stellar-orbit experiments around Sgr A$^*$ to compact objects and X-ray emitting stars at smaller angular separations. The GRAVITY collaboration has demonstrated that precision astrometry of individual stellar orbits can probe relativistic effects in the potential of the galactic center. Most recently, GRAVITY+ discovered S301, a star on a highly eccentric ($e\sim0.982$) 8.7 yr orbit that approaches Sgr A$^*$ close enough to become sensitive to frame dragging from the black hole spin through Lense-Thirring precession \cite{2026arXiv260712664E}. A high-resolution X-ray imager could provide an analogous experiment using bright X-ray sources whose optical or infrared counterparts may be too faint or confused for conventional astrometry. Monitoring multiple sources on short-period, highly-eccentric orbits can improve constraints on gravity under the most extreme environments, and open a new window into independent measurements of the spin of Sgr A$^*$ \cite{2026arXiv260724931P}.

\subsection{X-ray Parallaxes and Geometric Distances} 
\label{sec:parallax_distance}

Distances remain one of the dominant uncertainties in translating observed X-ray fluxes into intrinsic source properties and population statistics. Gaia provides exquisite parallaxes for optically accessible systems, while radio interferometry has measured geometric distances to a subset of radio-bright X-ray binaries and pulsars. However, many X-ray-selected sources lie in highly extincted or crowded regions, have intrinsically faint companions, or lack sufficiently bright radio counterparts. High-precision X-ray astrometry would provide direct geometric distances to these otherwise inaccessible populations.

The parallax of a source at distance $d$ is $\pi = 1/d_{\rm kpc}$ mas, corresponding to 200~$\mu$as at 5~kpc, 100~$\mu$as at 10~kpc, and 50~$\mu$as at 20~kpc. Sub-100-$\mu$as astrometry could therefore provide distance measurements for X-ray sources across much of the Milky Way, including sources in the Galactic bulge and toward the Galactic Center. These measurements would directly improve X-ray luminosity functions, Galactic spatial distributions, accretion-rate estimates, and compact-object parameter measurements, while providing the distance information needed for applications ranging from natal-kick studies to pulsar timing arrays.
    
\subsection{Astrometry in Globular Clusters} 

It has long been known that globular clusters host an overabundance of X-ray binaries compared to the Galactic disk. However, it is often difficult to identify the true optical/IR counterpart to the X-ray source due to the large chance coincidence probabilities (see e.g., \cite{2018ApJ...865...33H,2018MNRAS.479.2834H}). This is due to the large density of the optical/IR sources and the relatively large sizes of the Chandra positional uncertainties, which are often $\approx0.3''$ or larger (at 2$\sigma$), leading to large chance coincidence probabilities. An X-ray observatory with an angular resolution of 10 $m$as or better would measure the X-ray source positions very accurately (assuming accurate absolute astrometry), drastically reducing the chance coincidence probability, and allowing for a confident determination of the true optical/IR counterpart (or lack thereof) to the X-ray sources. Identifying the true counterpart of the X-ray sources is critical for source classification, particularly at the faint end of the X-ray luminosity function, where sources often have too few counts to constrain their spectra. The optical/IR counterpart can then be placed on the cluster color magnitude diagram (CMD) and its location on the CMD can be used to help determine the nature of the source (see e.g., \cite{2005ApJ...625..796H}). The precise determination of the counterpart and its photometric properties can also help to better understand the physical processes occurring in these systems. For example, locating the counterparts to X-ray emitting binary millisecond pulsars (MSPs) discovered in radio, known as spider binaries, can help better understand how MSPs ablate their companions as a function of spin-down power ($\dot{E}$), orbital separation, companion type, and system age (see e.g., \cite{2025ApJ...994....8K}). This is particularly important for ultracompact X-ray binaries, which are efficiently produced through dynamical interactions in dense stellar environments and can have extremely faint optical companions, making precise X-ray localization essential for identifying their counterparts and distinguishing them from the much more numerous CV population.

In addition to accurate counterpart determination, an angular resolution of $<1$ $m$as would allow for the measurement of X-ray proper motions. Several prominent clusters (e.g., 47 Tuc, Omega Cen, Terzan 5; \cite{2018A&A...612A.115N,2018ApJ...854...45L,2015ApJ...810...69M}) have proper motions of several $m$as yr$^{-1}$, which would easily be measurable with a $\sim1$ $m$as angular resolution instrument. This would determine cluster membership based solely on the X-ray proper motion, which would be useful for sources without counterparts (e.g., isolated MSPs) or those with counterparts too faint to be detected (e.g., black widow binaries with planet mass companions, or background active galactic nuclei). This would allow for a complete census of X-ray source populations to be determined down to some luminosity limit (determined by the detector sensitivity), which would be useful for direct comparison to simulations (see e.g., \cite{2026enap....3..458K} for a recent review). Furthermore, the internal motions of X-ray sources could be measured. For instance, Omega cen has a velocity dispersion of about 20 km $\rm s^{-1}$ (other clusters have typical values between 10-20 km $\rm s^{-1}$ \cite{2019ApJ...875....1M,2019MNRAS.482.5138B}), which at a distance of 5.4 kpc, would be measurable across a $\sim5$ year baseline \cite{2025ApJ...983...95H}. This would enable studies of the kinematics of X-ray emitting compact objects within globular clusters and comparison to simulations. Additionally, this would enable searches of fast moving X-ray sources, similar to those recently discovered in HST observations of Omega Cen \cite{2024Natur.631..285H}, which were used to place constraints on the location and mass of a potential intermediate mass black hole in the cluster.

\subsection{Refining Pulsar Timing Arrays with Pulsar Distance Estimates}
Pulsar timing arrays \cite{2025Ap&SS.370..124T} use timing residuals of pulsars to measure gravitational waves passing by both the Earth and the pulsar being used to trace the gravitational waves \cite{2011MNRAS.414.3251L}. The Hellings-Downs curve \cite{1983ApJ...265L..39H} measures the correlations between the pulsar timing residuals as a function of angular separation on the sky. This is useful at the present time because, without precise distances to the pulsars, the terms at the pulsar are not readily interpretable. The Earth terms for pulsars along similar lines of sight should be similar, and hence correlations and anticorrelations will build up with time and the number of tracers.

On the other hand, if several pulsars have sufficiently precise distance estimates, then the correlations between their timing residuals from their pulsar terms become meaningful, and can even be used to localize individual sources of gravitational waves, rather than just the stochastic background\cite{2011MNRAS.414.3251L}.

The challenge is that the precision of the distance measurement must be a fraction of the wavelength of the gravitational waves. For nanohertz waves, this means that ideally distances would be obtained to an accuracy of about 0.1 pc. Many of the NANOGrav pulsars are at distances of a few hundred pc. Taking 300 pc as a distance estimate, the parallax signal will be about 3.3 $m$as. With $\mu$as resolution, this would then mean that even a modest signal to noise measurement of the pulsar parallax would yield a sufficiently good distance estimate, but this can also be achieved through a combination of lower resolution and higher signal-to-noise. Astrometric precision at this scale is likely fundamentally impossible with ground-based radio telescopes because phase calibration of the ionosphere and troposphere can be done only via reference sources.

A complementary approach is provided by gamma-ray pulsar timing arrays (GPTAs), which use the rotational stability of $\gamma$-ray millisecond pulsars to search for the same nanohertz gravitational-wave signals targeted by radio PTAs.  The Fermi Large Area Telescope has already enabled timing of $\sim100$ millisecond pulsars in $\gamma$ rays, while future GeV observatories could increase this population to $10^3$--$10^4$ sources and achieve gravitational-wave sensitivities comparable to or exceeding those of current radio PTAs (\citealt{2026ApJ..1001L..36K}).  Gamma-ray timing is particularly complementary to radio measurements because it is unaffected by interstellar dispersion and scattering, suffers from fewer propagation-induced timing systematics, and accesses a somewhat different pulsar population.  Precise X-ray astrometric distances to $\gamma$-ray MSPs would therefore improve the interpretation of pulsar terms and strengthen joint radio--$\gamma$-ray PTA analyses, particularly for radio-faint or radio-quiet MSPs lacking other direct geometric distance measurements.

\section{Ideal Candidate Sources and Resolution Milestones}

The science return of an ultra-high-resolution X-ray observatory will depend strongly on the angular resolution that can be achieved, as well as on collecting area, field of view, astrometric stability, spectral response, timing capability, and imaging dynamic range.  This is particularly important for a pathfinder mission, for which the collecting area may be modest and photon statistics may limit the complexity of images that can be reconstructed. The most favorable early targets are therefore not necessarily those requiring the smallest physical scales, but rather bright sources for which a new angular-resolution regime enables a clear novel measurement: resolving previously identified extended structures on finer physical scales, measuring motion or elongation, separating a bright point source from surrounding emission, or determining a source centroid with precision substantially better than the nominal angular resolution. Table \ref{tab:ideal_candidates} summarizes particularly compelling science cases as a function of angular-resolution capability.

\begin{table*}
\centering
\caption{Representative science cases and candidate Galactic targets as a function of achievable angular resolution. The quoted fluxes and requirements are intended as order-of-magnitude guides for evaluating pathfinder concepts; in many cases, astrometric precision, effective area, spectral and timing resolution, dynamic range, field of view, count-rate capability, or target-of-opportunity response may be as important as the nominal angular resolution.}
\label{tab:ideal_candidates}
\small
% \vspace{-0.2cm}
\rowcolors{2}{white}{gray!20}
\resizebox{\textwidth}{!}{
\begin{tabular}{L{2.2cm} L{2.5cm} L{4.0cm} L{3.5cm} L{4.2cm} L{1.5cm}}
\hline
\textbf{Resolution} &
\textbf{Candidate source} &
\textbf{Primary science case} &
\textbf{Brightness / photon considerations} &
\textbf{Other key requirements} &
\textbf{Section} \\
\hline

$\sim100$ $m$as &
Crab pulsar/PWN &
Resolve inner knots, wisps, and pulsar/PWN structure; identify sites of particle acceleration &
$\sim1$ Crab; exceptionally photon-rich persistent source &
High dynamic range; high count-rate capability with minimal pile-up; high timing resolution; FoV $\gtrsim10''$ &
\ref{sec:mag_psr_PeV}, \ref{sec:star_death_SNR} \\

$\sim10$--100 $m$as &
Cas A / Tycho / SN 1006 &
Resolve synchrotron filaments, ejecta knots, and small-scale shock structure &
High surface brightness, although photons are distributed over an extended source &
High effective area; spatially resolved spectroscopy; FoV $\gtrsim10''$ &
\ref{sec:part_accel_SNR}, \ref{sec:star_death_SNR} \\

$\sim50$--100 $m$as &
SGR 1806--20-like giant flare &
Resolve and track ejecta following a magnetar giant flare &
Transient; extremely bright during the flare, followed by rapidly fading ejecta &
Rapid ToO response; high effective area; FoV $\gtrsim10''$ &
\ref{sec:mag_psr_PeV} \\ \hline

$\sim1$--10 $m$as &
SN 1987A &
Resolve shock and ejecta structure during the SN-to-SNR transition &
Relatively photon-starved compared with bright Galactic sources &
High effective area; spatially resolved spectroscopy; FoV $\gtrsim2''$ &
\ref{sec:star_death_SNR} \\

$\sim10$ $m$as &
SS 433 &
Trace baryonic jets inward and localize Fe-line-emitting or reheated plasma &
Bright persistent X-ray jet source &
High spectral resolution at Fe K; high dynamic range &
\ref{sec:wind_jet} \\

$\sim10$ $m$as &
XTE J1550--564 / MAXI J1820+070-like transients &
Track relativistic X-ray ejecta from launch through interaction with the ISM &
Most favorable during bright outbursts &
Rapid ToO response; high-precision astrometry &
\ref{sec:mag_psr_PeV}, \ref{sec:wind_jet} \\ \hline

$\sim1$ $m$as &
PSR B1259--63 &
Resolve the changing pulsar/stellar-wind interaction region across the eccentric orbit &
$L_X\sim10^{32}$--$10^{33}$ erg s$^{-1}$; significantly more photon-starved than bright XRBs &
High effective area; high dynamic range &
\ref{sec:HMGBs} \\

$\sim1$ $m$as &
Sgr A$^*$ &
Resolve accretion/outflow structure between Bondi and horizon scales &
Quiescent Chandra rate $\sim5.2\times10^{-3}$ counts s$^{-1}$ &
Effective area $\sim3000$ cm$^2$ at 4--8 keV; FoV $\sim1''$; high timing and spectral resolution &
\ref{sec:sgrA_jets_out}, \ref{sec:accretion_feed}, \ref{sec:galactic_center} \\

$\sim1$ $m$as imaging; sub-mas centroiding &
Bright pulsars / neutron stars &
Proper motions, parallaxes, and natal-kick measurements &
Best suited to bright point sources, where centroiding uses the full source photon count &
High-precision astrometry; high count-rate capability with minimal pile-up &
\ref{sec:natal_kicks}, \ref{sec:parallax_distance} \\ \hline

$\sim100~\mu$as &
Cygnus X-1 &
Probe wind-fed accretion on scales comparable to the binary separation; measure orbital astrometry and parallax &
$F_{0.5-10\,{\rm keV}}\sim10^{-8}$ erg s$^{-1}$ cm$^{-2}$; persistently bright &
High-precision astrometry; high dynamic range; high count-rate capability with minimal pile-up &
\ref{sec:BH_NS_orbit}, \ref{sec:comp_bin_feed}, \ref{sec:parallax_distance} \\

$\sim100~\mu$as imaging; few-$\mu$as centroiding &
Sco X-1 &
Precision parallax and orbital astrometry &
$F_{0.3-10\,{\rm keV}}\sim2$--$4\times10^{-7}$ erg s$^{-1}$ cm$^{-2}$; among the brightest persistent X-ray sources &
High-precision astrometry; high count-rate capability with minimal pile-up &
\ref{sec:BH_NS_orbit}, \ref{sec:parallax_distance} \\

$\sim100~\mu$as &
Bright obscured XRBs with poorly known companions &
X-ray-only orbital astrometry and astrometric mass-function constraints &
Prefer the brightest persistent systems to maximize centroid precision &
High-precision astrometry; &
\ref{sec:BH_NS_orbit}, \ref{sec:parallax_distance} \\ \hline

$\sim10~\mu$as &
V404 Cyg in outburst &
Search for asymmetric or extended emission associated with disk winds, stream--disk interactions, ejecta, and circumbinary material &
Can become extremely bright during outburst; quiescent state is substantially less favorable &
High dynamic range; rapid ToO response; high count-rate capability with minimal pile-up; high timing resolution &
\ref{sec:comp_bin_feed} \\

$\sim10~\mu$as &
Cygnus X-1 &
Image substantially inside the binary separation and probe the transition from focused stellar wind to accretion flow &
$F_{0.5-10\,{\rm keV}}\sim10^{-8}$ erg s$^{-1}$ cm$^{-2}$; persistent &
High dynamic range; high-precision astrometry; high count-rate capability with minimal pile-up &
\ref{sec:comp_bin_feed} \\

$\sim10~\mu$as astrometry &
Long-period bright X-ray binaries &
Measure orbital wobble, inclination, and astrometric mass functions &
Long-period and high-count-rate systems are favored because $a\propto P^{2/3}$ &
High-precision astrometry; high count-rate capability with minimal pile-up &
\ref{sec:BH_NS_orbit} \\

\hline
\end{tabular}
}
\end{table*}

These resolution milestones illustrate that a useful pathfinder does not need to achieve the ultimate $\mu$as goal to produce unique science. A $\sim10$--100 $m$as mission could target bright PWN, magnetar ejecta, SNR shocks, SN 1987A, and relativistic jets; a $\sim1$ $m$as instrument would add binary wind-interaction regions and precision X-ray astrometry; and sub-$m$as capabilities would begin to probe compact-binary separations and orbital motion directly. For a photon-starved mission, the strongest targets are therefore those that concentrate many photons into a compact region while offering a clear spatial or astrometric signature. The Crab, SS 433, PSR B1259--63, Cygnus X-1, Sco X-1, and bright transient ejecta from magnetars or X-ray binaries form a particularly favorable set of pathfinder targets, while sensitivity-demanding applications such as quiescent Sgr A$^*$ imaging, faint Galactic-center population studies, and spatially resolved spectroscopy of low-surface-brightness structures are more naturally associated with the next generation of high angular resolution observatory.

\section{Conclusions}
Milli- to micro-arcsecond X-ray imaging will open a fundamentally new regime for Galactic astrophysics, transforming many sources that are currently unresolved into spatially resolved laboratories. Across the science cases considered here, increased angular resolution is transformative, enabling qualitatively new measurements of physical scales, motions, and source environments. Resolving shocks, jets, accretion flows, stellar winds, binary orbits, and crowded source populations will connect phenomena that are presently studied only indirectly through spectra, variability, or population statistics. The progression from $m$as to tens-of-$\mu$as resolution therefore enables access to increasingly fundamental physical scales, from the structure of shocks and outflows to the dimensions of compact binaries and the immediate environments of black holes.

Such observations will provide new insight into how particles are accelerated and how stars and compact objects return energy to their environments. Spatially resolving synchrotron filaments, shock precursors, pulsar-wind structures, and relativistic ejecta will identify where particles are accelerated and how they subsequently escape and cool. In supernovae and their remnants, resolving ejecta, circumstellar material, and evolving shock fronts will connect the structure of the remnant to the progenitor and explosion mechanism, while measurements of compact remnants and their motion will constrain the natal kicks imparted during stellar collapse. These measurements will be especially powerful in a multi-messenger context. Neutrino and $\gamma$-ray observations can identify systems capable of accelerating particles to extreme energies, while high-resolution X-ray imaging can localize the shocks, jets, and pulsar-wind structures in which that acceleration occurs. Similarly, gravitational-wave detections of compact systems can be complemented by X-ray measurements of source position, distance, motion, and local environment. Rapid coordination with radio, optical, infrared, $\gamma$-ray, neutrino, and gravitational-wave facilities will therefore allow energetic Galactic transients to be followed simultaneously across different messengers and physical scales.

At even smaller scales, high-resolution imaging and astrometry will change how accretion and compact binaries are studied. In nearby X-ray binaries, resolving structures comparable to the binary separation will directly probe how stellar winds or Roche-lobe overflow feed an accretion flow, while imaging moving ejecta will connect changes close to the compact object to kinetic feedback on much larger scales. Around Sgr~A$^*$, $m$as imaging could bridge the gap between the Bondi-scale gas measured by Chandra and the horizon-scale flow probed by the EHT, directly testing how matter is lost between these regimes. Combining such imaging with high-time-resolution measurements would add important temporal information. Variability associated with pulsations, quasi-periodic oscillations, eclipses, flares, accretion-state changes, or the launching of transient ejecta could be connected directly to the structures in which those signals originate. Repeated imaging would link short-timescale variability to morphological evolution over days to years. Precise X-ray astrometry will add a complementary capability: parallaxes, proper motions, and astrometric binary orbits could yield geometric distances, compact-object velocities and masses, and potentially provide new probes of relativistic orbital dynamics around Sgr~A$^*$. Such distance measurements would also strengthen gravitational-wave studies with both radio and $\gamma$-ray pulsar timing arrays by allowing the pulsar terms to be modeled more accurately. In addition, resolving faint extended emission around bright X-ray binaries may place stringent requirements on imaging dynamic range, particularly for long-period systems and accretion-disk-corona sources in which scattered emission may arise over a broad range of radii. Detecting structures many orders of magnitude fainter than a bright and variable central source could therefore make dynamic range and temporal stability as important as nominal angular resolution.

Finally, this capability will transform our view of Galactic X-ray-source populations. Current surveys are often limited not by the existence of detectable X-ray photons, but by confusion, uncertain multiwavelength associations, and the inability to determine which physical source classes make up faint populations in environments such as the Galactic Center and globular clusters. Accurate X-ray positions and proper motions will permit secure counterpart identification, source classification, and membership determination, enabling population studies that are inaccessible today. Realizing this science will require more than angular resolution alone: stable and well-calibrated astrometry, sufficient collecting area for spatially resolved spectroscopy, useful spectral resolution, high-time-resolution capabilities that can connect variability to resolved structures, rapid response to transient and multi-messenger events, and high imaging dynamic range will all be essential complements. An observatory combining these capabilities with milli- to micro-arcsecond imaging will make spatial information a new fundamental dimension of X-ray astronomy and provide a uniquely powerful view of how compact objects form, accrete, accelerate particles, and evolve throughout the Galaxy.

\small
\bibliographystyle{aasjournalv7}
\bibliography{references}

\end{document}